\documentclass{article}
\usepackage[utf8]{inputenc} % allow utf-8 input
\usepackage[T1]{fontenc}    % use 8-bit T1 fonts
\usepackage[hidelinks]{hyperref}       % hyperlinks
\usepackage{url}            % simple URL typesetting
\usepackage{booktabs} % For better-looking tables
\usepackage{siunitx}        % professional-quality tables
\usepackage{amsfonts}       % blackboard math symbols
\usepackage{nicefrac}       % compact symbols for 1/2, etc.
\usepackage{microtype}   
\usepackage{orcidlink}

\usepackage[square, comma, numbers, sort&compress]{natbib}
\usepackage{arxiv}  

\usepackage{algorithm}
\usepackage{algpseudocode}
\usepackage{amssymb,amsbsy, amsmath}
\usepackage{graphicx}% Include figure files
\usepackage{rotating}
\usepackage[export]{adjustbox}
\usepackage{float}
\usepackage{array} 

\usepackage{dcolumn}% eqnarray table columns on decimal point
\usepackage[normalem]{ulem} %temporary
\usepackage{xr}
\usepackage{nameref}
\usepackage[version=4]{mhchem}
\definecolor{cadmiumgreen}{rgb}{0.0, 0.42, 0.24}
\usepackage{zref-xr}
\usepackage{booktabs} % For better looking tables
\usepackage{longtable} % For tables that span multiple pages
\usepackage[framemethod=TikZ]{mdframed}
\usepackage{arydshln}
\definecolor{customcolor}{HTML}{fffdf0}
\usepackage{tabularx}
\usepackage{authblk}
\usepackage{orcidlink}

\zxrsetup{toltxlabel}

\usepackage{setspace}
\usepackage{fullwidth}

\begin{document}

\title{Kernel learning assisted synthesis\\ condition exploration for ternary spinel}

\author[1]{\orcidlink{0000-0003-4737-7123}~Yutong Liu\textsuperscript{$\dagger$}}
\author[2]{\orcidlink{0000-0001-5696-9193}~Mehrad Ansari\textsuperscript{$\dagger$,$\ast$}}
\author[3]{\orcidlink{0009-0004-3383-0307}~Robert Black}
\author[1,2]{\orcidlink{0000-0003-2937-3188}~Jason Hattrick-Simpers\textsuperscript{$\ast$}}

\affil[1]{\small{Department of Materials Science and Engineering, University of Toronto, Toronto, ON, Canada}}
\affil[2]{\small{Acceleration Consortium, University of Toronto, Toronto, ON, Canada}}
\affil[3]{\small{Clean Energy Innovation Research Center, National Research Council Canada, Mississauga, ON, Canada}}

\renewcommand{\thefootnote}{\fnsymbol{footnote}}
\footnotetext{$\dagger$ These authors contributed equally.}
\footnotetext{$\ast$ Correspondence to: \{mehrad.ansari, jason.hattrick.simpers\}@utoronto.ca}

\maketitle

\begin{abstract}
\doublespacing
% Machine learning and high-throughput experimentation have accelerated the discovery of mixed metal oxide catalysts by leveraging their compositional flexibility. However, the lack of specified synthesis routes for certain structures remains a challenge in inorganic chemistry such as mixed metal oxide materials. Interpretable machine learning model is crucial and demanding, as it provides insights into the key factors governing phase formation. Here, we use kernel learning and SHAP values to assess the impact of individual experimental conditions on the formation of single-phase Fe$_2$(ZnCo)O$_4$, synthesized via high-throughput co-precipitation method. Contributions to single-phase spinel formation from SHAP analysis of precursor and precipitating agent align with crystal growth theories as expected. The amount of reagents

% These findings emphasize the need for interpretable ML approaches to refine synthesis methods and address environmental impacts in material design.

Machine learning and high-throughput experimentation have greatly accelerated the discovery of mixed metal oxide catalysts by leveraging their compositional flexibility. However, the lack of established synthesis routes for solid-state materials remains a significant challenge in inorganic chemistry. An \emph{interpretable} machine learning model is therefore essential, as it provides insights into the key factors governing phase formation. Here, we focus on the formation of single-phase Fe$_2$(ZnCo)O$_4$, synthesized via a high-throughput co-precipitation method. We combined a kernel classification model with a novel application of \emph{global} SHAP analysis to pinpoint the experimental features most critical to single phase synthesizability by interpreting the contributions of each feature.  \emph{Global} SHAP analysis reveals that precursor and precipitating agent contributions to single-phase spinel formation align closely with established crystal growth theories. These results not only underscore the importance of \emph{interpretable} machine learning in refining synthesis protocols but also establish a framework for data-informed experimental design in inorganic synthesis. 

\end{abstract}

\newpage

\doublespacing % change the line distance
\section{Introduction}
Mixed metal oxides (MMOs) exhibit excellent catalytic activity in reactions such as the oxygen evolution reaction~\cite{xu2021rational,tyndall2023understanding, li2022balair} and hydrogen evolution reaction~\cite{faid2021nicu,choi2021boosting}. 
MMOs can be composed of a wide selection of metal ions, ranging from earth-abundant alkali and transition metals to noble metals~\cite{gu2020oxygen,gawande2012role}. 
typically in various oxidation states, allowing for tunable structures and properties for specific applications. 
However, this flexibility of the design comes with challenges, as the vast number of possible combinations makes it difficult to identify compositions that show both good thermodynamic stability and can be successfully synthesized. 
To address this, machine learning (ML) or artificial intelligence (AI) has been frequently coupled with high-throughput experiments (HTE) to accelerate the discovery and optimization of targeted materials~\cite{zhou2024machine,jia2024machine}. 
With ML and HTE, scientists have not only made significant advances in automated experimentation, but also in autonomous experimentation, utilized to accelerate several aspects of material discovery~\cite{chang_efficient_2020,macleod_self-driving_2020,lu2024automated, zeni2025generative}.
Previous work has demonstrated significantly acceleration to the down-selecting of potential material candidates with enhanced properties, leading to more efficient experimentation and exploration compared to traditional means~\cite{pyzer-knapp_accelerating_2022,maqsood_future_2024}. 

% However, there is an evident gap between theory-based ML and real-life experiments. 
% Studies have highlighted even ML models that achieve strong performance on a single metric for test data often face challenges in generalizability, raising doubts about their applicability to other systems~\cite{steinmann2023machine,sutton2020identifying}. 
% First, models trained on specific datasets may perform well within the same or very similar material space, but often struggle when applied to new compounds or structures not represented in the training data~\cite{jain2024machine,bartel2020critical}.
% For our applications, this becomes problematic when the desire is to explore multiple combinations and permutations of MMOs, even those with similar structures.

Building a universally applicable model for synthesizability is challenging due to the limited dataset and the broad variability in synthesis methods. 
The limited data available often restricts the ability of machine learning models to generalize across different material spaces~\cite{steinmann2023machine,sutton2020identifying}, especially when dealing with new compounds or structures not covered in the training data~\cite{jain2024machine,bartel2020critical}. 
Consequently, it can be more practical to develop localized models that guide experimentalists in specific systems. 
Therefore, the most effective approach for synthesizing a particular material could be to create a surrogate model anchored in local experimental data, which raises a second challenge: the dataset would still be small and imbalanced, even with the use of automated high-throughput experiments to generate it~\cite{choubisa2020crystal}. Moreover, while ML can recommend promising materials with targeted compositions or structures, the synthesis route often remains unspecified for experimentalists~\cite{chen2024navigating,mannan2024navigating}. The core issue lies in the fact that researchers tend to rely on theoretical phase diagrams, leaving little guidance for designing practical, feasible routes to single-phase compounds~\cite{jansen2002concept,sun2016thermodynamic}.

This paper aims to tackle the challenges above by integrating kernel methods and explainable AI with real-world experiments, enabling the handling of small, imbalanced datasets while also guiding the development of practical, feasible synthesis routines.
Traditionally, theoretical and ML approaches have been developed primarily for problems with assumed linear settings~\cite{hofmann2008kernel}.
Few examples include calibration in spectroscopy~\cite{workman2018review}, predictive control and optimization~\cite{qin2014data}, and estimation of coefficient of thermal expansion~\cite{chew2024designing}.
In practice, however, these methods may not be applicable to complex real-world chemical systems, where the relationships between the process variables are non-linear~\cite{nelles2001nonlinear}. 
Kernel methods~\cite{scholkopf2002learning, shawe2004kernel, muller2001introduction,soentpiet1999advances,czekaj2005kernel} solve the nonlinearity problem by using a simple linear transformation manner.
The key idea is to project the sparse data onto a higher-dimensional space, where linear methods are more applicable~\cite{cover1965geometrical} and chances of over-fitting are less likely~\cite{cao2011exploring}. 
Kernel methods are performed in two successive steps: First, the training data in the input space is mapped onto a higher dimensional feature space, where sometimes even unknown features are induced by the kernel~\cite{mika1999fisher}.
In the second step, a linear method is applied to find a linear relationship in that feature space in a regression or a classification setting. Since everything is formulated in terms of kernel-evaluations, there is no need for any explicit calculations in the high-dimensional feature space~\cite{braun2008relevant}.

Herein, we leverage kernel learning and the SHapley Additive exPlanations (SHAP) to interpret the influence of synthesis conditions for the single-phase formation of a ternary spinel system Fe$_2$(ZnCo)O$_4$. 
Specifically, all samples are synthesized using a Chemspeed automation platform for better reproducibility and precious parameter control across the synthesis space. 
A kernel classification model is trained with the sparse independent experimental conditions including reagent concentrations, the amount of reagents, the reagent addition rate, and reagent addition order as features for single-phase synthesis. 
Through this, we aim to generate a synthesizability model.
With the synthesizability model, we introduce a \emph{global} SHAP analysis to interpret the positive-negative contributions of each feature for single-phase predictions. 
This novel application of SHAP allows for rapid in-silico navigation of the experimental parameter space, as well as identifying critical conditions to spinel synthesizability.
Surprisingly, our results indicate that the amount of reagents plays an important role in the high-throughput co-precipitation synthesis route.
This hints that the preferred conditions for material synthesis can cause environmental hazards by generating a certain amount of waste, especially for industrial production. 

This manuscript is organized as follows: Section~\ref{sec:results} presents the experimental results and exploration of the AI-suggested synthesis space. 
Section~\ref{sec:conclusion} follows with a discussion on the implications of our findings.
Finally, Section~\ref{sec:methods} provides details of our methodology, including the HTE synthesis and XRD measurement, kernel learning model, and performance evaluation metrics.

% \subsection{Advantages and Limitations}
\section{Results and Discussion}
\label{sec:results}

\begin{figure}[!t]
\includegraphics[width=\textwidth, center]{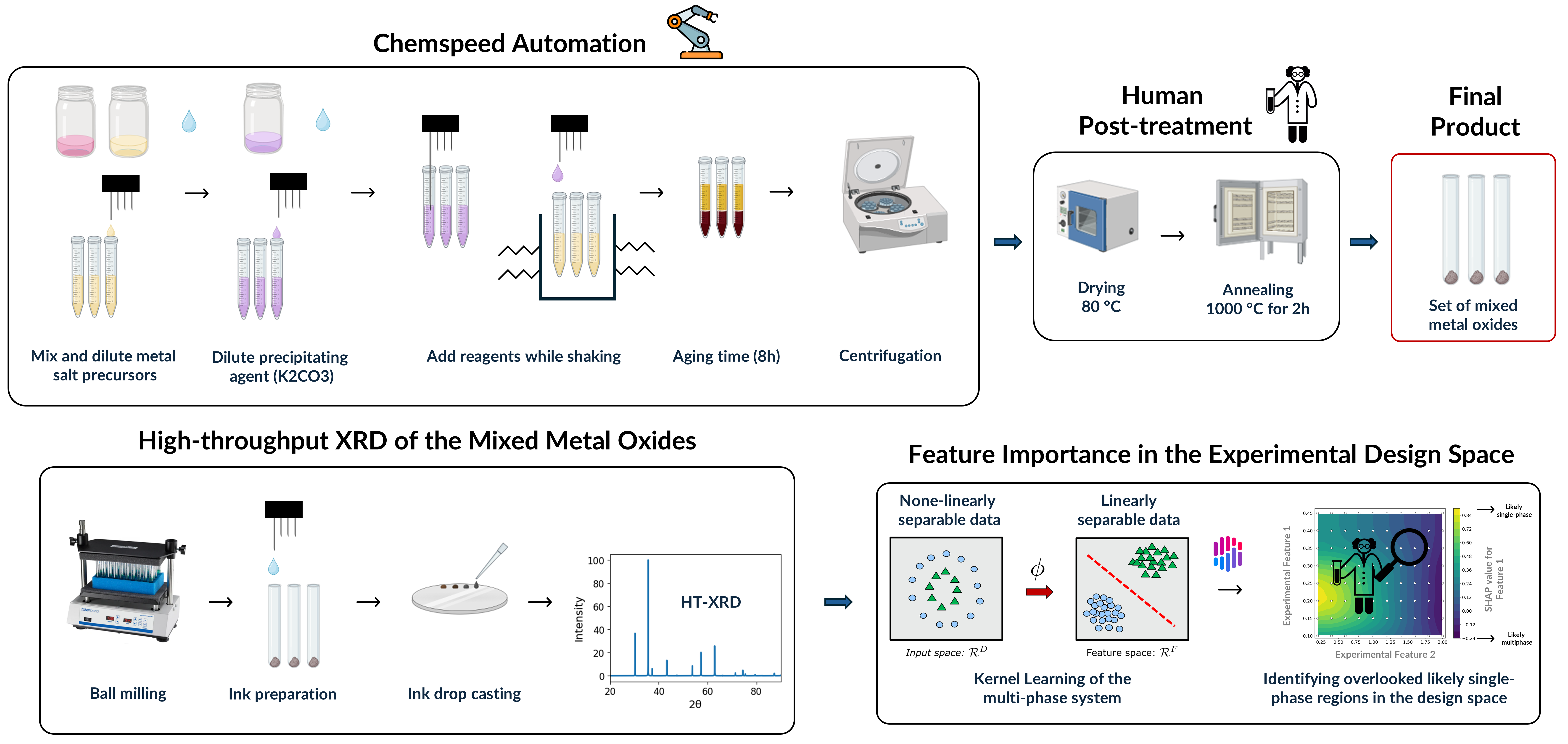}
\caption{A summary of the overall workflow from experiment to analysis. Fe$_2$(ZnCo)O$_4$ spinel samples are prepared by HT-synthesis using co-precipitation method on Chemspeed automated platform with a series of post-treatment. The final product is suspended in ink for HT-XRD measurement. The XRD results in single-phase classification are used as training data for the synthesizability model.}
\label{fig:workflow}
\end{figure}

\subsection{Synthesis method and parameter control}
\begin{figure}[!t]
\includegraphics[width=\textwidth, center]{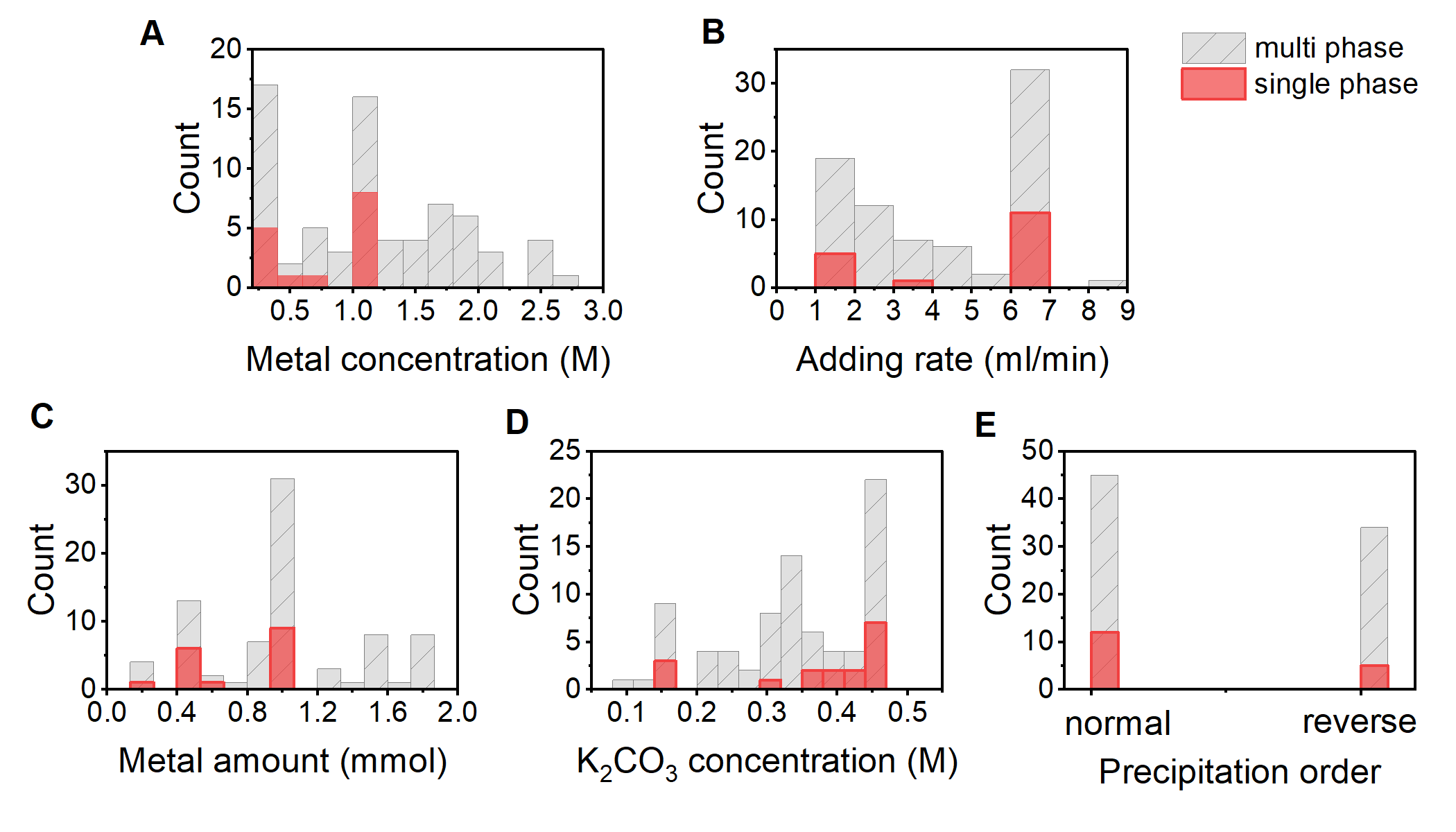}
\caption{Ground-truth distributions for single-phase versus multiple phase of five key experimental parameters in Fe$_2$(ZnCo)O$_4$ spinel
samples generated via Chemspeed automated platform. Phase labels are determined via high-throughput XRD.  (A) metal precursor concentration, (B) adding rate, (C) metal amount, (D) K$_2$CO$_3$ concentration, and (E) precipitation order. The total number of experiment samples is 70, with only 17 resulting in the desired single-phase solution.}
\label{fig:histrogram}
\end{figure}

Figure \ref{fig:workflow} demonstrates the workflow from synthesis and characterization to Kernal-method modeling and SHAP analysis. 
The targeted Fe$_2$(ZnCo)O$_4$ spinel powders were prepared by co-precipitation method using the Chemspeed platform (Swing XL model).
The detailed synthesis and parameter control are listed in Section~\ref{sec:synthesis}. Initially, a comprehensive list of experimental features was developed on the basis of the mechanism for crystal growth process, with strong consideration of pH, ionic strength, and precipitant quotients.
To mitigate multicollinearity and reduce chances of overfitting  — especially given the sparse experimental data — five key synthesis parameters were selected via feature importance (see Table~\ref{tab:synthesis_param}).
This was done by calculating the absolute correlation matrix for the generated experimental dataset using the Pearson's correlation coefficient. 
A predefined threshold of 0.55 was then used to identify pairs of highly linearly correlated features. 
For each pair with a correlation coefficient exceeding the threshold, one of the features was marked for removal to reduce redundancy and ensure feature independence.
The K$_2$CO$_3$ concentration and the metal precursor concentration represent the precursor concentrations prior to the precipitation reaction. 
Given the correlation between pH and K$_2$CO$_3$ concentration, pH was disregarded as an input feature.
All solutions are prepared from an initial stock solution, diluted accordingly to reach the desired concentrations. 
The metal precursor amount is calculated beforehand and imported into the platform for control. With these three parameters, all the necessary information for each precursor solution can be determined, including the volume of the highest concentrated precursors and the volume of water for dilution. 
The adding rate can be regulated by Chemspeed, and precipitation order is handled by experimental protocol design.
It should be noted here that the \emph{normal} precipitation order refers to adding the precipitating agents into metal precursor, while the other order is defined as \emph{reverse} precipitation order.
Figure~\ref{fig:histrogram} shows the distribution of each parameter from the conducted experiments.

High-throughput XRD results are used to label ground-truth data as single-phase or multi-phase. For our definition of single-phase,  we refer to a majority spinel phase as only observed in the XRD pattern.  
However, it is still possible that there is an undetectable secondary phase in the structure, given the limitations of our instrument. The other phase (or phases) can be one or a combination of Fe$_2$O$_3$, ZnO, or CoO metal oxide phases (Figures 
 S1).  
 This sparse dataset is significantly skewed toward the undesired multiphase solution, making it essential to employ kernel methods to assess how each feature positively contributes to achieving a single-phase solution.
% The iterations of experiments were selected in the region where uncertainty from the synthesizability kernel learning model is high for single-phase formation. 
\begin{table}[!t]

    \centering
    \caption{The five synthesis parameters controlled in the experiments. This selection aimed to mitigate multicollinearity and model overfitting in sparse data by applying a 0.55 threshold to the absolute Pearson correlation matrix to eliminate redundant features\label{tab:synthesis_param}}
    \begin{tabular}{p{0.3\textwidth}p{0.6\textwidth}}
        \toprule
        \textbf{Parameters} & \textbf{Description} \\ \midrule
        K$_2$CO$_3$ concentration (M) & K$_2$CO$_3$ concentration before mixing with metal precursors \\
        metal concentration (M) & All three metal precursors’ total concentration before mixing with K$_2$CO$_3$ solutions\\
        metal amount (mmol) & The amount of total metal precursors  \\
        adding rate (ml/min) & The adding rate of reagent A into B \\
        precipitation order & 0 for normal precipitation; 1 for reverse precipitation  \\ \bottomrule
    \end{tabular}
\end{table}

\subsection{Kernel classification of the ternary spinel}
A kernel learning model is used to classify the phase of the binary spinel, as described in method section~\ref{sec:ml}. 
By leveraging kernel-based learning approaches, the proposed classifier constructs a more refined and flexible decision boundary within the latent feature space. 
% As illustrated in Figure~\ref{fig:confmatrix}A, a preliminary step involves examining the absolute correlation matrix of the experimental features using Pearson’s correlation coefficient. Any feature pair with a correlation coefficient above 0.55 (e.g., precursor volumes, pH, and ionic strength) is considered highly correlated, and thus excluded from the selected features.
\begin{figure}[!b]
\includegraphics[width=\textwidth, center]{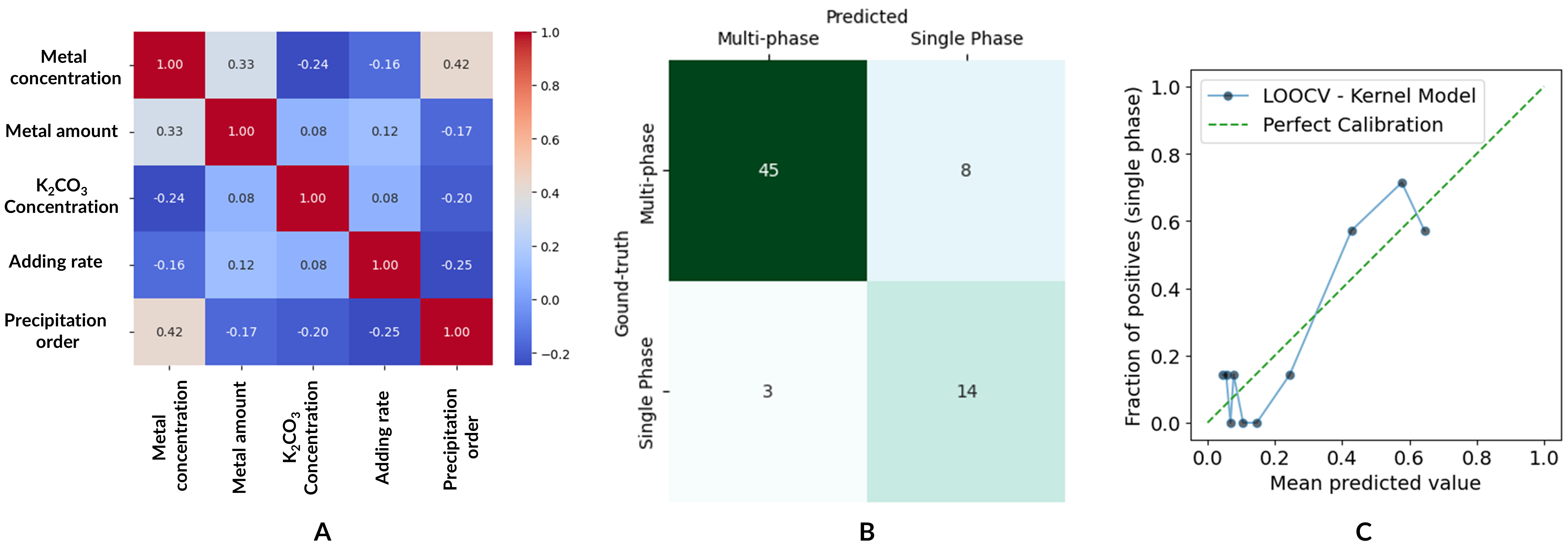}
\caption{Kernel learning is used in the ternary spinel synthesizability classification model. A) Absolute correlation matrix of the selected experimental features using Pearson’s correlation coefficient.
Other features with correlation $>=0.55$ are considered highly correlated, thus, disregarded.
B) Confusion matrix for the binary classification of the solution's phase. The leave-one-out cross-validation (LOOCV) accuracy and AUC are, 0.843 and 0.836, respectively.
C) The calibration curve shows that the model's predicted probabilities align with observed frequencies of single-phase outcomes, with some natural zigzagging due to sparse bin populations in LOOCV.)}

\label{fig:confmatrix}
\end{figure}
The kernel transformation not only captures subtle variations among data points but also helps mitigate overfitting, which is especially crucial given our limited sample size (70) and the heavy class imbalance skewed towards the undesired negative class (multiphase solution). 
Under these conditions, the kernel method helps preserve the integrity of the minority class representations while ensuring robust generalization and improved classification performance. 
This generalization is assessed via leave-one-out cross-validation (LOOCV), as shown in the confusion matrix in Figure~\ref{fig:confmatrix}B.
Notably, the LOOCV accuracy and the area under the curve (AUC) are 0.843 and 0.836, respectively, demonstrating the high efficacy of the classifier for synthesizability prediction.
Specifically, model evaluation on test data shows only 8 false negatives and 3 false positives, showing the robustness of the model in identifying actual positive (single-phase) experiments with a relatively high recall (0.824).
Figure~\ref{fig:confmatrix}C presents the calibration curve (reliability diagram), demonstrating how the predicted probabilities of the model align with the observed frequency of the single phase outcomes. 
The ``zigzagging'' trend, rather than a perfectly smooth line, is primarily due to having very few samples in each bin as a result of LOOCV.
 Despite this, the overall trend clearly demonstrates that higher predicted probabilities consistently correspond to higher proportions of actual single-phase experiments, confirming the model's ability to meaningfully rank experimental outcomes.
 The lack of predictions above 0.7 reflects the classifier’s tendency toward moderate probability estimates in a constrained data setting. 
Rather than assigning near‐certain predictions, the model remains conservative, which reduces the upper range of predicted probabilities. 
This conservative approach is preferable for our application, as it ensures caution and reliability when dealing with inherently uncertain data.

We capitalize on the model's high recall to systematically explore the design space for the desired single-phase solutions. To guide this exploration, we employ an uncertainty sampling acquisition policy that prioritizes experiments based on the model’s confidence. 
Specifically, we generate a comprehensive set of 43,000 candidate experiments by taking the Cartesian product of the relevant set of 5 experimental parameters (see Section~\ref{sec: methods-shap} in Methods). 
The uncertainty of each candidate is quantified as the absolute difference from a probability of 0.5, with higher values representing predictions closer to the decision boundary, enabling us to focus on parameter combinations where the model is most certain. 
This strategy ensures that limited experimental resources are directed toward the most informative samples, accelerating the discovery and validation of stable single-phase regions.

Additionally, we leverage SHAP (SHapley Additive exPlanations) values to gain deeper insights into each parameter’s positive contribution toward achieving a single-phase solution.
This interpretative step is a novel component of our approach, offering valuable guidance on which parameters most strongly promote single-phase formation, thus, further refining our exploration strategy for identifying stable single-phase regions.

\subsection{Model interpretability and feature contribution in the synthetic design space}

We use SHAP values to measure the contribution of each feature by comparing the classifier’s prediction with and without that feature, then averaging these contributions over all possible coalitions in the synthetic design space. 
This process yields both \emph{local} and \emph{global} interpretations of model behavior. While the \emph{local} interpretation reveals how a particular feature influences the prediction for an individual data instance (see Figure S4 in SI), the \emph{global} interpretation identifies which features have the greatest overall impact on the model’s predictions across all generated synthetic experiments~\cite{lundberg2017unified,lundberg2020local}.

Formally, for each feature \(i\) and data instance \(x\), the \emph{local} SHAP value \(\phi_i(f; x)\) is defined as:
\begin{equation}
\phi_i(f; x) 
= \sum_{S \subseteq N \setminus \{i\}}
\frac{|S|!\,(|N| - |S| - 1)!}{|N|!}
\Bigl[
f_x(S \cup \{i\}) - f_x(S)
\Bigr],
\end{equation}
where \(N\) is the full set of features, \(S \subseteq N \setminus \{i\}\) is a subset of features excluding \(i\), and \(f_x(S)\) is the model’s prediction for instance \(x\) when using only features in \(S\).

Meanwhile, the \emph{global} SHAP value \(\overline{\phi}_i(f)\) for feature \(i\) is obtained by averaging the \emph{local} SHAP values over all samples \(x \in \mathcal{D}\) in the synthetic design space:
\begin{equation}
\overline{\phi}_i(f) 
= \frac{1}{|\mathcal{D}|} \sum_{x \in \mathcal{D}} \phi_i(f; x),
\end{equation}
where \(\mathcal{D}\) is the complete set of generated samples and \(|\mathcal{D}|\) is the total number of samples in that space.

\begin{figure}[!b]
    \includegraphics[width=\textwidth, center]{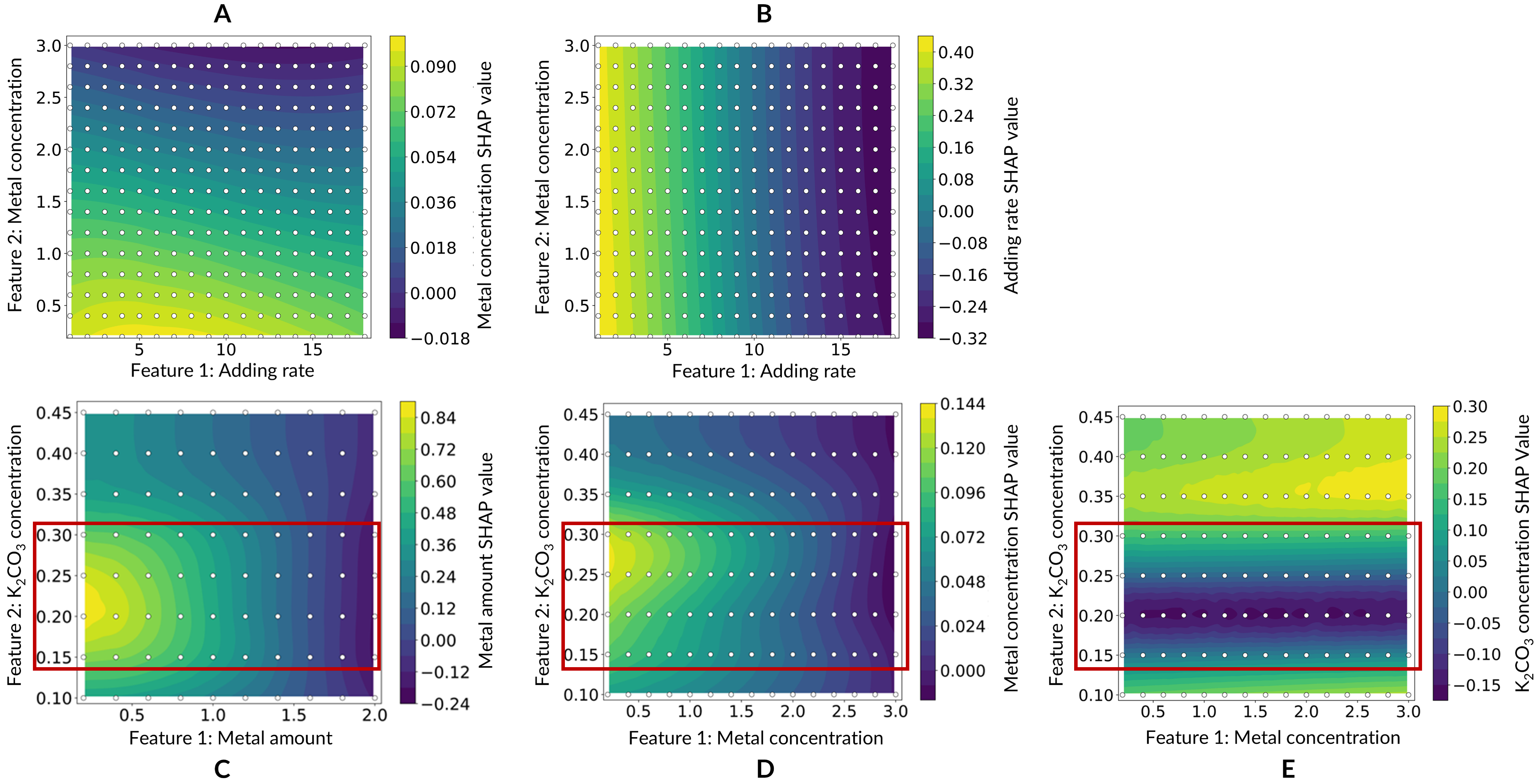}
    \caption{Contour plots of the \emph{global} SHAP values for different pairs of experimental features, aggregated over the synthetic space of 43,000 experiment samples. The phase of the binary spinel is inferred by the kernel classifier, and the global SHAP value offers a comprehensive view of how each feature combination positively influences single-phase formation in alignment with theoretical and experimental expectations.
    Within the K$_2$CO$_3$ concentration range of 0.15--0.3\,M, all samples exhibit at least one secondary phase, creating a \emph{missing region} (marked with red) with very few single-phase outcomes (panel C).
    Despite additional experiments confirming single-phase synthesis is possible, the overall success rate in this region remains significantly lower than elsewhere, underscoring the distinct effect of K$_2$CO$_3$ on phase formation.}
    \label{fig:shap}
\end{figure}

Figure~\ref{fig:shap} illustrates \emph{global} SHAP values for individual features as contour plots spanning the full synthetic design space. Although SHAP values are computed \emph{locally} for each instance, once a contour map covering the entire range of two features is produced, it effectively aggregates these values across all samples, providing a \emph{global} perspective.
Here, we provide a detailed discussion on the pairwise contour plots of the \emph{global} SHAP value for each feature in the synthetic space, where the phase of the binary spinel is inferred by our kernel model. 
This analysis offers insights into how these interpretative results align with existing theoretical understanding and experimental expectations. These plots ultimately reveal key regions in the synthetic design space, where certain parameter combinations most strongly favor single-phase formation.

\subsubsection*{Adding rate and metal concentration}
Global SHAP values for adding rate and metal precursor concentration align with chemists' intuition. As shown in Figure~\ref{fig:shap}A and B, in the projection of metal concentration and adding rate, both parameters prefer smaller values (yellow regions) to form single-phase spinel materials,  favouring steady and stable reaction conditions. 
In contrast, higher metal precursor concentrations and adding rate preferably result in multiple-phase existence of the final precipitants. 
Figure~\ref{fig:shap}A shows that the SHAP value of metal precursor concentration has the highest positive contribution in the region where metal precursor concentration is lower than 0.5 M and adding rate between 2 to 8 ml/min. In the top region of Figure~\ref{fig:shap}A, where metal concentration is above 2 M, the metal concentration feature exhibits a negative effect on the targeted phase generation. This result aligns with the experimental observation (Figure~\ref{fig:histrogram}) that only multiple phases form at metal concentration higher than 1.5 M. Here an adding rate of 3 ml/min is equal to a dropwise every second, which is commonly used in traditional co-precipitation method. However, the contribution of adding rate is quite independent compared to other parameters. The contour plot (Figure~\ref{fig:shap}B) of adding rate contribution shows clear steps from 1 to 8 ml/min. Furthermore, the quantitative comparison between those two features’ contribution should not be neglected.
The highest positive SHAP value from adding rate (0.4) and metal concentration (0.09) have different orders of magnitude, indicating the direct role of adding rate in the formation of single phases.

\subsubsection*{The amount of reagents}
The amounts of reagents have a significant impact on the single-phase formation as observed in Figure~\ref{fig:shap}C. 
The amounts of reagents in this analysis refers to the metal amount parameter, as K$_2$CO$_3$ amount is consistently set to 1.5 times the metal amount to ensure complete reaction and is exclued from the feature correlation analysis~\cite{biesuz2018synthesis}. 
We should note that the metal amount and metal concentration are two distinct features. 
 The metal amount represents the total metal cations in each sample, not the concentration of the prepared metal salt solution. 
Figure \ref{fig:shap}C shows the metal amount contribution contour profile in the metal amount versus K$_2$CO$_3$ concentration dimension. The highest SHAP value of 0.85 indicates the pivotal role of reagent amounts, especially compared with the highest SHAP value 0.144 from metal concentration. See (Figure~\ref{fig:shap}D), where metal concentration SHAP values are plotted against the same secondary parameter K$_2$CO$_3$ concentration. The highest SHAP value from metal precursor amount in Figure~\ref{fig:shap}C suggest this parameter is crucial in the region where the K$_2$CO$_3$ concentration stays between 0.18 to 0.25 M and the metal amount is lower than 0.3 mmol. However, the distribution from experimental data (Figure~\ref{fig:histrogram}) indicates there is no single-phase generated when K$_2$CO$_3$ concentration is in the range from 0.15 to 0.3 M. The influence of K$_2$CO$_3$ concentration will be further discussed in the next section. The amounts of reagents are usually not considered in traditional experiments, but SHAP analysis results identify its important role in high-throughput synthesis. 
The sample volume and reaction scale for high-throughput synthesis are usually limited by the modular and protocol design.
Studies have shown that different reaction scales can lead to different products in Suzuki-Miyaura reaction~\cite{jurica2021automation}. Considering the stochastic nucleation process, it is understandable that the nucleation induction probability depends on the reactor scale~\cite{sun2017induction}.

\subsubsection*{K$_2$CO$_3$ concentration}
\begin{table}[!t]
    \centering
    \caption{The possible reactions during precipitant growth process.\label{tab:k2co3} }
    \begin{tabular}{p{0.5\textwidth}>{\centering\arraybackslash}p{0.5\textwidth}}
        \toprule
        \textbf{Reaction type} & \textbf{Reaction equations} \\
        \midrule
        Potassium carbonate dissociation & 
        \begin{minipage}[c]{\linewidth}
            \centering
            \vspace{0pt}
            \begin{equation*}
                \text{K}_2\text{CO}_3 \rightarrow 2\text{K}^+ + \text{CO}_3^{2-}   
            \end{equation*}
            \begin{equation*}
                \text{CO}_3^{2-} + \text{H}_2\text{O} \leftrightharpoons \text{HCO}_3^- + \text{OH}^-
            \end{equation*}
            \begin{equation*}
                \text{HCO}_3^- + \text{H}_2\text{O} \leftrightharpoons \text{H}_2\text{CO}_3 + \text{OH}^-
            \end{equation*}
            \begin{equation*}
                \text{H}_2\text{CO}_3 \leftrightharpoons \text{CO}_2 + \text{H}_2\text{O}
            \end{equation*}
        \end{minipage} \\
        \midrule
        Precipitant growth & 
        \begin{minipage}[c]{\linewidth}
            \centering
            \vspace{0pt}
            \begin{equation*}
                \text{M}^{n+} + \frac{n}{2}\text{CO}_3^{2-} \leftrightharpoons \text{M}(\text{CO}_3)_{\frac{n}{2}}
            \end{equation*}
            \begin{equation*}
                \text{M}^{n+} + n\text{OH}^- \leftrightharpoons \text{M}(\text{OH})_n 
            \end{equation*}
        \end{minipage} \\
        \midrule
        Precipitant dissolution & 
        \begin{minipage}[c]{\linewidth}
            \centering
            \vspace{0pt}
            \begin{equation*}
                \text{M}(\text{CO}_3)_{\frac{n}{2}} \leftrightharpoons \text{M}^{n+} + \frac{n}{2}\text{CO}_3^{2-} 
            \end{equation*}
            \begin{equation*}
                \text{M}(\text{OH})_n  \leftrightharpoons \text{M}^{n+} + n\text{OH}^-
            \end{equation*}
        \end{minipage} \\
        \bottomrule
    \end{tabular}
\end{table}

\begin{figure}[!t]
\includegraphics[width=\textwidth, center]{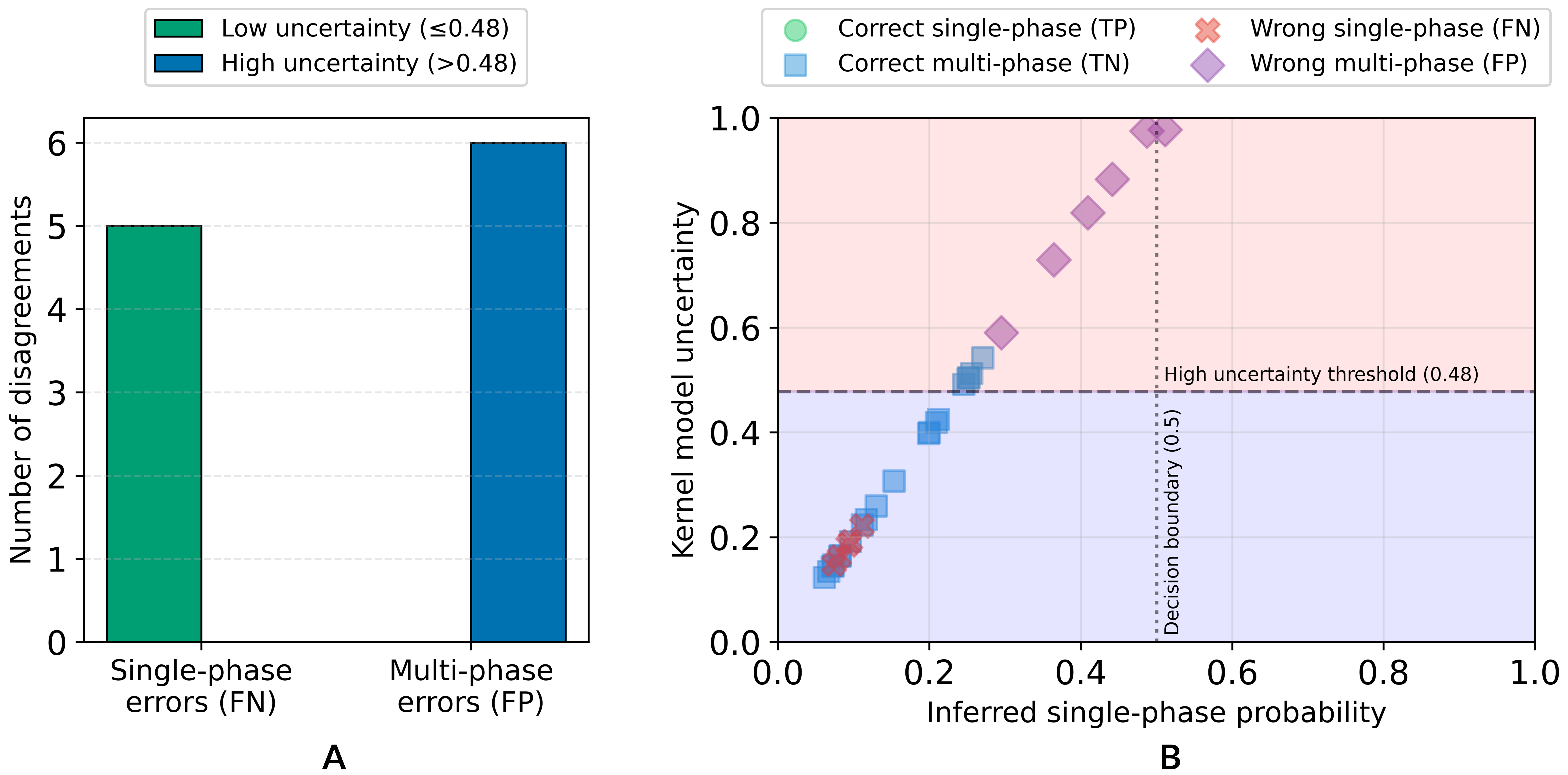}
\caption{\label{fig:calibrarion} Kernel model calibration assessment. The high uncertainty threshold was set at the 66th percentile (0.47). A) Distribution of kernel model's disagreements (errors) in the missing region by uncertainty level and class.
B) Probability vs uncertainty of all samples in the missing region. 100\% of the multi-phase errors (FP) occur in the high uncertainty region, while most correct predictions (TP/TN) cluster in the low uncertainty regions with higher confidence (probabilities further from 0.5), demonstrating appropriate model calibration.
}
\end{figure}

We use K$_2$CO$_3$ as the precipitating agent in the co-precipitation synthesis method~\cite{biesuz2018synthesis}.
Within the range of 0.15 to 0.3 M concentration, all 12 samples have at least one secondary phase according to XRD phase identification. 
Therefore, we call this region a \emph{missing region} due to the low probability of a successful single-phase synthesis. 
The distribution of other parameters in this missing region are shown in Figure S2 in SI. The distributions of other parameters in the missing region fall within the same range as the rest of the data (Figure~\ref{fig:histrogram}), allowing us to rule out improper condition selection as the cause.
We conducted an additional 31 experiments in the missing region to verify this observation (Figure S3).
Although only 5 additional single-phase samples are fabricated in this region, the single-phase ratio (9.1\%) is still more than two times lower than the other regions (26.9\%).
We used these additional experiments to assess the calibration of our kernel model (Figure~\ref{fig:calibrarion}).
Of the 31 predictions made by the kernel model, 20 aligned with the ground truth, while 11 did not.
Figure~\ref{fig:calibrarion}A shows the distribution of these disagreements (errors) across uncertainty levels, separated by class type. Uncertainty was calculated as the distance from 0.5 probability, with higher values indicating predictions closer to the decision boundary. We set the high uncertainty threshold at the 66th percentile (0.47).
Overall, 54.5\% of model errors occurred in regions of high uncertainty, suggesting reasonable calibration. For multi-phase samples, all disagreements (100\%) occurred in high uncertainty regions, indicating the model appropriately expresses low confidence when it misclassifies multi-phase samples as single-phase.
In contrast, single-phase misclassifications showed lower uncertainty, suggesting an area for potential model improvement, especially by the addition of more single-phase samples, thus, reducing the class imbalance in the training data.
Figure~\ref{fig:calibrarion} B visualizes the model's predictions of the new experiments from the missing region in the probability-uncertainty space.
It is observed that most multi-phase errors (purple diamonds) cluster in the high uncertainty region near the decision boundary, while correct predictions (green circles and blue squares) generally exhibit higher confidence (probabilities much further from 0.5). 
The observed pattern confirms that the model's expressed uncertainty correlates with error likelihood, particularly for multi-phase samples.

The additional results confirm our initial assumption regarding the interesting influence of K$_2$CO$_3$ in co-precipitation synthesis. Figure~\ref{fig:shap}E clearly shows that K$_2$CO$_3$ concentration in the missing region has only negative contributions regardless of the metal precursor concentration. As a weak base, K$_2$CO$_3$ can slow down the reaction, enabling more kinetic control, similar to the usage of a strong base NaOH with a complexation agent ammonia to synthesize dense and spherical hydroxides, as demonstrated for Li-ion battery cathode materials~\cite{mugumya2022synthesis}.
K$_2$CO$_3$ can provide hydroxide ion source and also buffer pH to ensure consistent pH. 
However, it has been reported that metal hydroxide particle growth occurs with dissolution even in a continuous stirred tank reactor (CSTR)~\cite{van2009analysis,wang2011growth,mugumya2022synthesis}. 
Each crystallization and dissolution for different metal cations has a different equilibrium constant, generating a complex net of reactions.
For K$_2$CO$_3$ as the precipitating agent, the net reaction includes the dissociation of potassium carbonate, hydrolysis of carbonate, followed by bicarbonate, and the simultaneous growth and dissolution of precipitants. (Table~\ref{tab:k2co3}). 
Here $\text{M}^\text{n+}$ represents all metal cations in both A and B sites ($\text{Fe}^\text{3+}$, $\text{Co}^\text{2+}$ and $\text{Zn}^\text{2+}$).
It is reasonable to assume that with the presence of K$_2$CO$_3$ at a certain pH range favors a secondary phase metal hydroxide formation since the precipitations start to form at different pH for each metal cation species~\cite{mugumya2022synthesis}.

% \begin{table}[!t]
%     \centering
%     \caption{The possible reactions during precipitant growth process.}
%     \label{tab:k2co3}
%     \begin{tabular}{l|l} 
    
%     \hline
%     \rule{0pt}{15pt} 
%         \textbf{ Reaction type}&  \textbf{Reaction equation}\\ 
%         \hline
        
%          Potassium carbonate dissociation&  \begin{equation}
%  \text{K}_2\text{CO}_3 \rightarrow 2\text{K}^+ + \text{CO}_3^{2-}   
% \end{equation}\\ 
%          &  \begin{equation}
%     \text{CO}_3^{2-} + \text{H}_2\text{O} \leftrightharpoons \text{HCO}_3^- + \text{OH}^-
% \end{equation}\\ 
%          &  \begin{equation}
%     \text{HCO}_3^- + \text{H}_2\text{O} \leftrightharpoons \text{H}_2\text{CO}_3 + \text{OH}^-
% \end{equation}\\ 
%          &  \begin{equation}
%     \text{H}_2\text{CO}_3 \leftrightharpoons \text{CO}_2 + \text{H}_2\text{O}
% \end{equation}\\ 
% \hline
%          Precipitant growth&  \begin{equation}
% \text{M}^{n+} + \frac{n}{2}\text{CO}_3^{2-} \leftrightharpoons \text{M}(\text{CO}_3)_{\frac{n}{2}}
% \end{equation}\\ 
%          &  \begin{equation}
% \text{M}^{n+} + n\text{OH}^- \leftrightharpoons \text{M}(\text{OH})_n 
% \end{equation}\\ 
% \hline
%          Precipitant dissolution&  \begin{equation}
%  \text{M}(\text{CO}_3)_{\frac{n}{2}} \leftrightharpoons \text{M}^{n+} + \frac{n}{2}\text{CO}_3^{2-} 
% \end{equation}\\ 
%          &  \begin{equation}
% \text{M}(\text{OH})_n  \leftrightharpoons \text{M}^{n+} + n\text{OH}^-
% \end{equation}\\ 
% \hline
%     \end{tabular}
  
% \end{table}

\subsection{Crystal nucleation and growth}
Findings from the global SHAP analysis can be corroborated through the underlying thermodynamics and kinetics of crystal growth theory. 
Crystal growth theory has been established since Burton, Cabrera and Frank proposed their basic theory (i.e. BCF theory)~\cite{burton1951growth,uwaha2016introduction}. 
BCF theory is widely applied in crystal growth studies from vapor, solution and melt systems, and provides a microscopic description on the mechanism of crystal nucleation, growth, and precipitation. 
For solution-based synthesis, the precipitate or crystal growth can be separated into two stages: nucleation and growth. 
The first step, nucleation, occurs rapidly as an induced supersaturation of the solution. Metastable polymorphs and metaphases form rapidly during this process due to the non-equilibrium state, and are dependent on the supersaturation ratio governed by the addition rate of K$_2$CO$_3$ and concentration of metal hydroxides. Our conclusions from the SHAP analysis aligns with this theory, with slower addition of K$_2$CO$_3$ resulting in a higher probability of single-phase growth, in agreement with slow K$_2$CO$_3$ addition resulting in a low supersaturation ratio favoring fewer nucleation sites. The second stage is the growth process. After the precipitate forms, metal cations in the solution are continuously absorbed onto the nuclei while some ions diffuse from the surface of nuclei, reaching a dynamic equilibrium. One metric used to describe the interaction between the particle and supernatant is zeta potential $\zeta$, which depends on pH, ionic strength and ion properties~\cite{duan2014zeta}. For single-phase materials, a high zeta potential is preferred for controlled growth states for the increased repulsive electrostatic interactions~\cite{adair2001surface}. SHAP analysis results align with the role of zeta potential in the crystal growth process. The missing region in the middle range of K$_2$CO$_3$ concentration may result from a low zeta potential since both low and high pH values can significantly increase the zeta potential (Figure~\ref{fig:shap}E)~\cite{serrano2023zeta}. Similarly, lower precursor concentration with lower ionic strength can prompt a thicker double layer around the particles and lead to higher zeta potential, making the overall crystal nucleation and growth process easier to control. This explains the highest SHAP values only appear in the low metal concentration region (Figure~\ref{fig:shap}D).

We have discussed how each experimental parameter contributes to the prediction results in above SHAP analysis. All the parameters are related to the thermodynamics and kinetics in the precipitate formation process. Our SHAP value analysis qualitatively aligns with expectations of BCF theory. However, our SHAP analysis also reveals that there is a high degree of complexity. From the results, we can conclude that the complexity of mixed metal oxide synthesis systems rise as the number of metal cations and competing metastable phases increase. Computational studies including machine learning have made it possible to identify and downselect promising materials, but the difficulty of synthesizing the predicted composition or structure still exists in real-life laboratories. Therefore, we believe that surrogate synthesizability models focusing on material nucleation and growth should be developed with effective and generalized descriptors. Additionally, due to the inherent nature and randomness of crystal growth, we propose that synthesizability should be modeled as a probabilistic outcome. This study provides a preliminary inspiration for the synthesizability model pipeline for the community. 

\subsection{Limitations in high-throughput synthesis}
During the synthesis exploration process, we notice several limitations and challenges in high throughput synthesis that need attention from us and the broader community. Questions including the flexibility of experiments, reaction considerations and chemical considerations have been thoroughly discussed previously~\cite{christensen2021automation}. Here, we use our high-throughput synthesis as an example to discuss additional layers to this question. 

The reaction scale depends on the reactor size from given automation platform. It limits both the parameters we can change and the range within which we can adjust them. In our experiments, the maximum total volume of two precursor solutions is limited to 13 ml with 2 ml extra volume reaction space, since the modules in our Chemspeed platform require the usage of 15 ml tubes in each well. The total volume constrains the grid search space of the experimental parameters, in both the search parameter number and range  of each parameter. The theoretical parameter space should have six dimensions including the two reagents’ concentrations and amounts (from which their volumes can be calculated), adding rate and precipitation order. 
The total volume limitation reduces one dimension in the parameter space. Therefore, we ultimately use a searching space with five dimensions. The range of volume also restricts other parameters’ searching range. If the reagent concentrations are too diluted, their volume can easily go over the total volume limit of our experimentation. Also, the volume limit regulates the amount of final products, which can cause trouble for continuous characterization measurement.

Our results further demonstrate that the reagent amount proves to be an important feature in resulting the final product phase  (Figure~\ref{fig:shap}). As discussed in the previous section, it is understandable that a smaller reaction amount can favor one phase formation in a fixed reactor space. When the nucleation process is considered as a stochastic event, the probability of forming nuclei can be modeled by a Poisson process, which depends on the volume of available space~\cite{kashchiev1991induction}. The high-throughput synthesis reactor can be considered as a closed system with a fixed amount of reagents. Theoretical studies have shown that crystallization in a closed system would go through a bulk metastable phase~\cite{sun2017induction}. Compared with open systems in inorganic chemistry synthesis, a flow chemistry setup like a continuous stirred tank reactor can be hard to achieve for high-throughput experiments. The fixed reactor size and closed system design constrain how much we can explore experimental conditions with high-throughput experimentation. 
\section{Conclusion}
\label{sec:conclusion}
In this study, Fe$_2$(ZnCo)O$_4$
  spinel samples were synthesized via a co-precipitation method on the Chemspeed automation platform, and their phases were verified using high-throughput X-ray diffraction.
  By systematically varying five experimental parameters to explore single-phase formation conditions, we integrated a kernel classification model with a novel application of \emph{global} SHAP analysis to pinpoint the experimental features most critical to synthesizability. 
  These findings reveal that reagent amounts exert a stronger influence than precursor concentrations and underscore the advantages of combining automated experimentation with interpretable machine learning. 
The global SHAP analysis is aligned with the classic BCF theory as K$_2$CO$_3$ plays an importance role in both crystal nucleation and growth stages by influencing phase stability and growth thermodynamics.
SHAP analysis highlights the complexity of single-phase formation amid competing phases.
  Our approach offers a promising guidance for high-throughput inorganic material synthesis, aligning both theoretical insights and practical expectations.

% In conclusion, co-precipitation method was performed using Chemspeed automation platform to prepare Fe$_2$(ZnCo)O$_4$ spinel samples. High-throughput X-ray diffraction technique was used to identify the phase of produced materials. Five experimental parameters were systematically varied within the range of single-phase spinel formation. A kernel  classification model was applied to assess synthesizability, complemented by SHAP analysis to evaluate the impact of individual features. The study reveals that the reagent amounts have a significant influence compared with precursor concentrations. This study provides a guideline for inorganic material synthesis in high-throughput experiment field. 

\section{Methods}
\label{sec:methods}

\subsection{HT synthesis}
\label{sec:synthesis}
Fe$_2$(ZnCo)O$_4$ spinel was synthesized using Chemspeed Swing XL platform with co-precipitation method as shown in Figure~\ref{fig:workflow}. Firstly, highly concentrated metal precursor stock solutions (2M) and precipitating agent K$_2$CO$_3$ (1M) were manually prepared and placed in Chemspeed platform. Metal precursors including ferric chloride (99+\%, Thermo Scientific), zinc nitrate (98\%, Sigma-Aldrich)and cobalt nitrate ($\geq$97.0\%, Sigma-Aldrich) were used and mixed with the fixed ratio of 4:1:1. Potassium carbonate (Certified ACS, Fisher Chemical) was used as a precipitating agent to adjust pH for the formation of mixed metal hydroxides/oxides. 
The amount of total metal precursor and the amount of K$_2$CO$_3$ were always kept at a ratio of 1:1.5~\cite{biesuz2018synthesis}. Secondly, both metal precursors and K$_2$CO$_3$ solutions were diluted to different concentrations respectively. The diluted solutions were drawn from the bottom of each tube and then dispensed into the tube for complete dilution.
Both reagents were also moved to shaking zone for a 5-minute shaking with 100 rpm. Thirdly, the precipitating agents (or metal precursors) were added at controlled rates into metal precursors (or precipitating agents) to form precipitants while shaking at 100 rpm. 
The precipitant-solution mixtures were shaking for another 10 min and then aging for 8h in the Chemspeed. Finally, the precipitant-solution mixed samples were centrifuged and rinsed with DI water for three cycles. All the above operations were carried out by Chemspeed platform with five parameters controlled (Table~\ref{tab:synthesis_param}). 
After automated synthesis, the powders were tried at $80^{\circ}$ in a vacuum oven and then transferred in a box furnace for 2h annealing at $1000^{\circ}$. After ball milling treatment, 0.006 ml nafion (5\%), 0.03 ml isopropanol and 0.06 ml water were added for each 1 mg fine powder to prepare inks for high-throughput XRD measurement. 

\subsection{HT-XRD measurement}
The inks in section~\ref{sec:synthesis} were drop cast on silicon wafers for high-throughput XRD measurement. X-ray diffraction (XRD) measurements were conducted using a Bruker D8 Discover diffractometer, equipped with a Cu microfocus X-ray source ($\lambda$ = 0.15418 nm) operating at 50 kV and 1000 $\mu$A with a 0.4 mm slit (Figure S1 in the supporting information). 
 XRD data were collected in the $2\theta$ range from 5$^\circ$ to 80$^\circ$, with a step size of 0.04$^\circ$ and a dwell time of 1.2 seconds per increment.
The diffraction patterns were recorded in 1D mode at ambient temperature, with the detector aperture set to 62 mm x 20 mm. The Si wafer with samples was placed on the measuring stage, and the coordinates of the first five samples were determined using the overhead camera integrated into the diffractometer.
A script was then employed to apply translation and 2D rotation to calculate the remaining samples' coordinates.

\subsection{Kernel learning}
\label{sec:ml}

With a kernel, data can be nonlinearly mapped from original input space $\mathcal{R}^{D}$ onto a feature space $\mathcal{R}^{F}$, with input and feature dimension $D$ and $F$, respectively.
For transformation $\phi:\mathcal{R}^{D} \rightarrow \mathcal{R}^{F}$, a kernel function is defined as
\begin{equation}
    \label{eq:kernel}
    K({\bf x},{\bf y}) = \langle \phi({\bf x}), \phi({\bf y})\rangle_F = \phi({\bf x})^T\phi({\bf y}),
\end{equation}
where $\langle\cdot\rangle_F$ is the inner product. 
This means that a kernel is required to work on scalar products of type ${\bf x}^T{\bf y}$ that can be translated into scalar products $\phi({\bf x})^T\phi({\bf y})$ in the feature space. 
On the other hand, as long as $F$ is an inner product space, the explicit representation of $\phi$ is not necessary and the kernel function $K({\bf x},{\bf y})$ can  be directly evaluated as~\cite{boser1992training}.
This is also known as the \emph{kernel trick}~\cite{soentpiet1999advances}, removing the need for  the explicit evaluation of the computationally expensive transformation $\phi$.
Interestingly, many algorithms for regression and classification can be reformulated in terms of the kernelized dual representation, where the kernel function arises naturally~\cite{muller2001introduction}. 
The transformation can be done implicitly by the choice of the kernel. 
In specific, the kernel encodes a real valued similarity between inputs (i.e. experimental conditions) {\bf x} and {\bf y}.
The similarity measure is defined by the representation of the system which is then used in combination with linear or non-linear kernel functions such as Gaussian, Laplace, polynomial and sigmoid kernels. 
Alternatively, the similarity measure can be encoded \emph{directly} into the kernel, leading to a wide variety of kernels in the chemical domain~\cite{schutt2020machine}.
In this setting, the defined binary kernel function needs to be non-negative, symmetric and point-separating (i.e. $
\langle x, x' \rangle = 0 \iff x = x'$).
Given the numerical experimental features, we define the distance \((d)\) between the parameter choices in experiments \(x\) and \(y\) as the \(l_2\) norm (Euclidean distance):

\begin{equation}
\label{eq:distance-kernel}
d(x, y) = \sqrt{\sum_{i=1}^{n} (x_i - y_i)^2}.
\end{equation}

Here, \(x_i\) and \(y_i\) represent the values of the \(i\)-th parameter in experiments \(x\) and \(y\), respectively, and \(n\) is the total number of experimental parameters.

%% Moved this to results sec
% \subsubsection{Feature importance and explainability}
% To address the multicollinearity among features, we calculated the absolute correlation matrix for the generated experimental dataset using the Pearson's correlation coefficient. 
% A predefined threshold of 0.55 was then used to identify pairs of highly linearly correlated features. 
% For each pair with a correlation exceeding the threshold, one of the features was marked for removal to reduce redundancy and ensure feature independence.

\subsubsection{Kernel-Based support vector classifier}\label{sec: kernel-classifier}

We used a support vector classifier (SVC) with a custom distance-based kernel (Equation~\ref{eq:distance-kernel}) to predict single-phase spinel formation, and evaluated its performance using leave-one-out cross-validation (LOOCV). 
At each LOOCV iteration, one sample experiment was reserved for testing, while the remaining samples formed the training set. We first scaled each feature into the range \([0,1]\) using a transformation. 
Then the pairwise Euclidean distances among the training samples (\(\mathbf{X}_{\mathrm{train}}\)) and between each test sample and the training set were calculated. 
These distance matrices replaced the original input features, allowing the SVC to learn a classification boundary based on the distance profiles in the higher dimensional space.
At inference, we obtained both a class label (single-phase vs.\ multi-phase) and the decision-function score. 
The latter is the absolute distance from the decision boundary, which was normalized to serve as an uncertainty measure.

\subsubsection{Synthetic design space exploration}
\label{sec: methods-shap}

We defined a parameter grid to systematically explore the experimental design space. 
The experimental variables include metal concentration, which takes values from 0.2 to 3.0 in increments of 0.2; metal amount, ranging from 0.2 to 2.0 with a step of 0.2; K$_2$CO$_3$ concentration, varying from 0.1 to 0.45 in steps of 0.05; rate, an integer parameter spanning from 1 to 18; and precipitation order, a binary parameter (0 or 1). 
The Cartesian product of these parameters yields 43,000 unique experimental conditions. 
For each sample point, our kernel classifier is applied to estimate the uncertainty, measured as the deviation from a probability of 0.5, where higher values signify predictions that are nearer to the decision boundary. 
This approach enables efficient in-silico navigation of the high-dimensional design space, while providing quantitative uncertainty estimates for the phase predictions.

\subsection{Performance metrics}

To evaluate the performance of the binary classifier, we use two common metrics: \emph{accuracy} and the \emph{area under the receiver operating characteristic} (AUROC).
Accuracy is calculated as the proportion of correct predictions (true positives and true negatives) out of the total number of predictions:
\begin{equation}
\mathrm{Accuracy} = \frac{TP + TN}{TP + TN + FP + FN},
\end{equation}
where \(TP\) is the number of true positives, \(TN\) is the number of true negatives, \(FP\) is the number of false positives, and \(FN\) is the number of false negatives.
AUROC measures the ability of the classifier to rank positive instances higher than negative ones, accounting for different discrimination thresholds. Formally, it is given by the integral of the true positive rate (\(TPR\)) as a function of the false positive rate (\(FPR\)):
\begin{equation}
\mathrm{AUROC} = \int_{0}^{1} \mathrm{TPR} \, d(\mathrm{FPR}),
\end{equation}
where \(\mathrm{TPR} = \frac{TP}{TP + FN}\) and \(\mathrm{FPR} = \frac{FP}{TN + FP}\).

\section*{Acknowledgements}
This research was undertaken thanks in part to funding provided to the University of Toronto's Acceleration Consortium from the Canada First Research Excellence Fund: Grant number - CFREF-2022-00042.
The authors also gratefully acknowledge the partial financial support from Materials for Clean Fuel (MCF) Challenge program at National Research Council of Canada (NRC).

% \section{Author declarations}
\subsection*{Conflict of interest}
The authors have no conflicts to disclose.
% \subsection{Author contributions}

\section*{Data and Code Availability}
All data  and code used to produce results in this study are publicly available in the following GitHub repository:
\href{https://github.com/AccelerationConsortium/gremlin}{\color{blue}{{https://github.com/AccelerationConsortium/gremlin}}}.

% \printbibliography

\bibliographystyle{unsrtnat}

\bibliography{bibliography}

\begin{thebibliography}{55}
\providecommand{\natexlab}[1]{#1}
\providecommand{\url}[1]{\texttt{#1}}
\expandafter\ifx\csname urlstyle\endcsname\relax
  \providecommand{\doi}[1]{doi: #1}\else
  \providecommand{\doi}{doi: \begingroup \urlstyle{rm}\Url}\fi

\bibitem[Xu et~al.(2021)Xu, Fan, Zou, Fu, Dong, Dou, Wang, Chen, Yin, Al-Mamun, et~al.]{xu2021rational}
Yiming Xu, Kaicai Fan, Yu~Zou, Huaiqin Fu, Mengyang Dong, Yuhai Dou, Yun Wang, Shan Chen, Huajie Yin, Mohammad Al-Mamun, et~al.
\newblock Rational design of metal oxide catalysts for electrocatalytic water splitting.
\newblock \emph{Nanoscale}, 13\penalty0 (48):\penalty0 20324--20353, 2021.

\bibitem[Tyndall et~al.(2023)Tyndall, Gannon, Hughes, Carolan, Pinilla, Ja{\'s}kaniec, Spurling, Ronan, McGuinness, McEvoy, et~al.]{tyndall2023understanding}
Daire Tyndall, Lee Gannon, Lucia Hughes, Julian Carolan, Sergio Pinilla, Sonia Ja{\'s}kaniec, Dahnan Spurling, Oskar Ronan, Cormac McGuinness, Niall McEvoy, et~al.
\newblock Understanding the effect of mxene in a tmo/mxene hybrid catalyst for the oxygen evolution reaction.
\newblock \emph{npj 2D Materials and Applications}, 7\penalty0 (1):\penalty0 15, 2023.

\bibitem[Li et~al.(2022)Li, Liu, Qin, and Liu]{li2022balair}
Haisen Li, Huihui Liu, Qing Qin, and Xien Liu.
\newblock Balair double mixed metal oxides as competitive catalysts for oxygen evolution electrocatalysis in acidic media.
\newblock \emph{Inorganic Chemistry Frontiers}, 9\penalty0 (4):\penalty0 702--708, 2022.

\bibitem[Faid et~al.(2021)Faid, Barnett, Seland, and Sunde]{faid2021nicu}
Alaa~Y Faid, Alejandro~Oyarce Barnett, Frode Seland, and Svein Sunde.
\newblock Nicu mixed metal oxide catalyst for alkaline hydrogen evolution in anion exchange membrane water electrolysis.
\newblock \emph{Electrochimica Acta}, 371:\penalty0 137837, 2021.

\bibitem[Choi et~al.(2021)Choi, Surendran, Kim, Lim, Lim, Park, Kim, Han, and Sim]{choi2021boosting}
Hyeonuk Choi, Subramani Surendran, Dohun Kim, Yoongu Lim, Jaehyoung Lim, Jihyun Park, Jung~Kyu Kim, Mi-Kyung Han, and Uk~Sim.
\newblock Boosting eco-friendly hydrogen generation by urea-assisted water electrolysis using spinel m 2 geo 4 (m= fe, co) as an active electrocatalyst.
\newblock \emph{Environmental Science: Nano}, 8\penalty0 (11):\penalty0 3110--3121, 2021.

\bibitem[Gu et~al.(2020)Gu, Camayang, Samira, and Nikolla]{gu2020oxygen}
Xiang-Kui Gu, John Carl~A Camayang, Samji Samira, and Eranda Nikolla.
\newblock Oxygen evolution electrocatalysis using mixed metal oxides under acidic conditions: Challenges and opportunities.
\newblock \emph{Journal of catalysis}, 388:\penalty0 130--140, 2020.

\bibitem[Gawande et~al.(2012)Gawande, Pandey, and Jayaram]{gawande2012role}
Manoj~B Gawande, Rajesh~K Pandey, and Radha~V Jayaram.
\newblock Role of mixed metal oxides in catalysis science—versatile applications in organic synthesis.
\newblock \emph{Catalysis Science \& Technology}, 2\penalty0 (6):\penalty0 1113--1125, 2012.

\bibitem[Zhou et~al.(2024)Zhou, Wang, Tang, Ling, Yu, and Chen]{zhou2024machine}
Pan Zhou, Ming Wang, Fei Tang, Liu Ling, Hongfang Yu, and Xi~Chen.
\newblock Machine learning accelerates the screening of efficient metal-oxide catalysts for photocatalytic water splitting.
\newblock \emph{Materials Research Bulletin}, page 112956, 2024.

\bibitem[Jia and Li(2024)]{jia2024machine}
Xue Jia and Hao Li.
\newblock Machine learning enabled exploration of multicomponent metal oxides for catalyzing oxygen reduction in alkaline media.
\newblock \emph{Journal of Materials Chemistry A}, 2024.

\bibitem[Chang et~al.()Chang, Nikolaev, Carpena-Núñez, Rao, Decker, Islam, Kim, Pitt, Myung, and Maruyama]{chang_efficient_2020}
Jorge Chang, Pavel Nikolaev, Jennifer Carpena-Núñez, Rahul Rao, Kevin Decker, Ahmad~E. Islam, Jiseob Kim, Mark~A. Pitt, Jay~I. Myung, and Benji Maruyama.
\newblock Efficient closed-loop maximization of carbon nanotube growth rate using bayesian optimization.
\newblock 10\penalty0 (1):\penalty0 9040.
\newblock ISSN 2045-2322.
\newblock \doi{10.1038/s41598-020-64397-3}.

\bibitem[{MacLeod} et~al.(){MacLeod}, Parlane, Morrissey, Häse, Roch, Dettelbach, Moreira, Yunker, Rooney, Deeth, Lai, Ng, Situ, Zhang, Elliott, Haley, Dvorak, Aspuru-Guzik, Hein, and Berlinguette]{macleod_self-driving_2020}
B.~P. {MacLeod}, F.~G.~L. Parlane, T.~D. Morrissey, F.~Häse, L.~M. Roch, K.~E. Dettelbach, R.~Moreira, L.~P.~E. Yunker, M.~B. Rooney, J.~R. Deeth, V.~Lai, G.~J. Ng, H.~Situ, R.~H. Zhang, M.~S. Elliott, T.~H. Haley, D.~J. Dvorak, A.~Aspuru-Guzik, J.~E. Hein, and C.~P. Berlinguette.
\newblock Self-driving laboratory for accelerated discovery of thin-film materials.
\newblock 6\penalty0 (20):\penalty0 eaaz8867.
\newblock \doi{10.1126/sciadv.aaz8867}.

\bibitem[Lu et~al.(2024)Lu, Pan, Mo, and Fang]{lu2024automated}
Jia-Min Lu, Jian-Zhang Pan, Yi-Ming Mo, and Qun Fang.
\newblock Automated intelligent platforms for high-throughput chemical synthesis.
\newblock \emph{Artificial Intelligence Chemistry}, page 100057, 2024.

\bibitem[Zeni et~al.(2025)Zeni, Pinsler, Z{\"u}gner, Fowler, Horton, Fu, Wang, Shysheya, Crabb{\'e}, Ueda, et~al.]{zeni2025generative}
Claudio Zeni, Robert Pinsler, Daniel Z{\"u}gner, Andrew Fowler, Matthew Horton, Xiang Fu, Zilong Wang, Aliaksandra Shysheya, Jonathan Crabb{\'e}, Shoko Ueda, et~al.
\newblock A generative model for inorganic materials design.
\newblock \emph{Nature}, pages 1--3, 2025.

\bibitem[Pyzer-Knapp et~al.()Pyzer-Knapp, Pitera, Staar, Takeda, Laino, Sanders, Sexton, Smith, and Curioni]{pyzer-knapp_accelerating_2022}
Edward~O. Pyzer-Knapp, Jed~W. Pitera, Peter W.~J. Staar, Seiji Takeda, Teodoro Laino, Daniel~P. Sanders, James Sexton, John~R. Smith, and Alessandro Curioni.
\newblock Accelerating materials discovery using artificial intelligence, high performance computing and robotics.
\newblock 8\penalty0 (1):\penalty0 1--9.
\newblock ISSN 2057-3960.
\newblock \doi{10.1038/s41524-022-00765-z}.

\bibitem[Maqsood et~al.()Maqsood, Chen, and Jacobsson]{maqsood_future_2024}
Ayman Maqsood, Chen Chen, and T.~Jesper Jacobsson.
\newblock The future of material scientists in an age of artificial intelligence.
\newblock 11\penalty0 (19):\penalty0 2401401.
\newblock ISSN 2198-3844.
\newblock \doi{10.1002/advs.202401401}.

\bibitem[Steinmann et~al.(2023)Steinmann, Wang, and Seh]{steinmann2023machine}
Stephan~N Steinmann, Qing Wang, and Zhi~Wei Seh.
\newblock How machine learning can accelerate electrocatalysis discovery and optimization.
\newblock \emph{Materials Horizons}, 10\penalty0 (2):\penalty0 393--406, 2023.

\bibitem[Sutton et~al.(2020)Sutton, Boley, Ghiringhelli, Rupp, Vreeken, and Scheffler]{sutton2020identifying}
Christopher Sutton, Mario Boley, Luca~M Ghiringhelli, Matthias Rupp, Jilles Vreeken, and Matthias Scheffler.
\newblock Identifying domains of applicability of machine learning models for materials science.
\newblock \emph{Nature communications}, 11\penalty0 (1):\penalty0 4428, 2020.

\bibitem[Jain(2024)]{jain2024machine}
Anubhav Jain.
\newblock Machine learning in materials research: developments over the last decade and challenges for the future.
\newblock \emph{Current Opinion in Solid State \& Materials Science}, 2024.

\bibitem[Bartel et~al.(2020)Bartel, Trewartha, Wang, Dunn, Jain, and Ceder]{bartel2020critical}
Christopher~J Bartel, Amalie Trewartha, Qi~Wang, Alexander Dunn, Anubhav Jain, and Gerbrand Ceder.
\newblock A critical examination of compound stability predictions from machine-learned formation energies.
\newblock \emph{npj computational materials}, 6\penalty0 (1):\penalty0 97, 2020.

\bibitem[Choubisa et~al.(2020)Choubisa, Askerka, Ryczko, Voznyy, Mills, Tamblyn, and Sargent]{choubisa2020crystal}
Hitarth Choubisa, Mikhail Askerka, Kevin Ryczko, Oleksandr Voznyy, Kyle Mills, Isaac Tamblyn, and Edward~H Sargent.
\newblock Crystal site feature embedding enables exploration of large chemical spaces.
\newblock \emph{Matter}, 3\penalty0 (2):\penalty0 433--448, 2020.

\bibitem[Chen et~al.(2024)Chen, Cross, Miara, Cho, Wang, and Sun]{chen2024navigating}
Jiadong Chen, Samuel~R Cross, Lincoln~J Miara, Jeong-Ju Cho, Yan Wang, and Wenhao Sun.
\newblock Navigating phase diagram complexity to guide robotic inorganic materials synthesis.
\newblock \emph{Nature Synthesis}, pages 1--9, 2024.

\bibitem[Mannan et~al.(2024)Mannan, Bihani, Krishnan, and Mauro]{mannan2024navigating}
Sajid Mannan, Vaibhav Bihani, NM~Anoop Krishnan, and John~C Mauro.
\newblock Navigating energy landscapes for materials discovery: Integrating modeling, simulation, and machine learning.
\newblock \emph{Materials Genome Engineering Advances}, 2\penalty0 (1):\penalty0 e25, 2024.

\bibitem[Jansen(2002)]{jansen2002concept}
Martin Jansen.
\newblock A concept for synthesis planning in solid-state chemistry.
\newblock \emph{Angewandte Chemie International Edition}, 41\penalty0 (20):\penalty0 3746--3766, 2002.

\bibitem[Sun et~al.(2016)Sun, Dacek, Ong, Hautier, Jain, Richards, Gamst, Persson, and Ceder]{sun2016thermodynamic}
Wenhao Sun, Stephen~T Dacek, Shyue~Ping Ong, Geoffroy Hautier, Anubhav Jain, William~D Richards, Anthony~C Gamst, Kristin~A Persson, and Gerbrand Ceder.
\newblock The thermodynamic scale of inorganic crystalline metastability.
\newblock \emph{Science advances}, 2\penalty0 (11):\penalty0 e1600225, 2016.

\bibitem[Hofmann et~al.(2008)Hofmann, Sch{\"o}lkopf, and Smola]{hofmann2008kernel}
Thomas Hofmann, Bernhard Sch{\"o}lkopf, and Alexander~J Smola.
\newblock Kernel methods in machine learning.
\newblock \emph{The annals of statistics}, 36\penalty0 (3):\penalty0 1171--1220, 2008.

\bibitem[Workman~Jr(2018)]{workman2018review}
Jerome~J Workman~Jr.
\newblock A review of calibration transfer practices and instrument differences in spectroscopy.
\newblock \emph{Applied spectroscopy}, 72\penalty0 (3):\penalty0 340--365, 2018.

\bibitem[Qin et~al.(2014)Qin, Lysecky, and Sprinkle]{qin2014data}
Xiao Qin, Susan Lysecky, and Jonathan Sprinkle.
\newblock A data-driven linear approximation of hvac utilization for predictive control and optimization.
\newblock \emph{IEEE Transactions on Control Systems Technology}, 23\penalty0 (2):\penalty0 778--786, 2014.

\bibitem[Chew et~al.(2024)Chew, Afzal, Chandrasekaran, Kamps, and Ramakrishnan]{chew2024designing}
Alex~K Chew, Mohammad Atif~Faiz Afzal, Anand Chandrasekaran, Jan~Henk Kamps, and Vaidya Ramakrishnan.
\newblock Designing the next generation of polymers with machine learning and physics-based models.
\newblock \emph{Machine Learning: Science and Technology}, 5\penalty0 (4):\penalty0 045031, 2024.

\bibitem[Nelles(2001)]{nelles2001nonlinear}
Oliver Nelles.
\newblock Nonlinear dynamic system identification.
\newblock In \emph{Nonlinear System Identification}, pages 547--577. Springer, 2001.

\bibitem[Sch{\"o}lkopf et~al.(2002)Sch{\"o}lkopf, Smola, Bach, et~al.]{scholkopf2002learning}
Bernhard Sch{\"o}lkopf, Alexander~J Smola, Francis Bach, et~al.
\newblock \emph{Learning with kernels: support vector machines, regularization, optimization, and beyond}.
\newblock MIT press, 2002.

\bibitem[Shawe-Taylor et~al.(2004)Shawe-Taylor, Cristianini, et~al.]{shawe2004kernel}
John Shawe-Taylor, Nello Cristianini, et~al.
\newblock \emph{Kernel methods for pattern analysis}.
\newblock Cambridge university press, 2004.

\bibitem[Muller et~al.(2001)Muller, Mika, Ratsch, Tsuda, and Scholkopf]{muller2001introduction}
K-R Muller, Sebastian Mika, Gunnar Ratsch, Koji Tsuda, and Bernhard Scholkopf.
\newblock An introduction to kernel-based learning algorithms.
\newblock \emph{IEEE transactions on neural networks}, 12\penalty0 (2):\penalty0 181--201, 2001.

\bibitem[Soentpiet et~al.(1999)]{soentpiet1999advances}
Rosanna Soentpiet et~al.
\newblock \emph{Advances in kernel methods: support vector learning}.
\newblock MIT press, 1999.

\bibitem[Czekaj et~al.(2005)Czekaj, Wu, and Walczak]{czekaj2005kernel}
Tomasz Czekaj, Wen Wu, and Beata Walczak.
\newblock About kernel latent variable approaches and svm.
\newblock \emph{Journal of Chemometrics: A Journal of the Chemometrics Society}, 19\penalty0 (5-7):\penalty0 341--354, 2005.

\bibitem[Cover(1965)]{cover1965geometrical}
Thomas~M Cover.
\newblock Geometrical and statistical properties of systems of linear inequalities with applications in pattern recognition.
\newblock \emph{IEEE transactions on electronic computers}, \penalty0 (3):\penalty0 326--334, 1965.

\bibitem[Cao et~al.(2011)Cao, Liang, Xu, Hu, Zhang, and Fu]{cao2011exploring}
Dong-Sheng Cao, Yi-Zeng Liang, Qing-Song Xu, Qian-Nan Hu, Liang-Xiao Zhang, and Guang-Hui Fu.
\newblock Exploring nonlinear relationships in chemical data using kernel-based methods.
\newblock \emph{Chemometrics and Intelligent Laboratory Systems}, 107\penalty0 (1):\penalty0 106--115, 2011.

\bibitem[Mika et~al.(1999)Mika, Ratsch, Weston, Scholkopf, and Mullers]{mika1999fisher}
Sebastian Mika, Gunnar Ratsch, Jason Weston, Bernhard Scholkopf, and Klaus-Robert Mullers.
\newblock Fisher discriminant analysis with kernels.
\newblock In \emph{Neural networks for signal processing IX: Proceedings of the 1999 IEEE signal processing society workshop (cat. no. 98th8468)}, pages 41--48. Ieee, 1999.

\bibitem[Braun et~al.(2008)Braun, Buhmann, and M{\"u}ller]{braun2008relevant}
Mikio~L Braun, Joachim~M Buhmann, and Klaus-Robert M{\"u}ller.
\newblock On relevant dimensions in kernel feature spaces.
\newblock \emph{The Journal of Machine Learning Research}, 9:\penalty0 1875--1908, 2008.

\bibitem[Lundberg(2017)]{lundberg2017unified}
Scott Lundberg.
\newblock A unified approach to interpreting model predictions.
\newblock \emph{arXiv preprint arXiv:1705.07874}, 2017.

\bibitem[Lundberg et~al.(2020)Lundberg, Erion, Chen, DeGrave, Prutkin, Nair, Katz, Himmelfarb, Bansal, and Lee]{lundberg2020local}
Scott~M Lundberg, Gabriel Erion, Hugh Chen, Alex DeGrave, Jordan~M Prutkin, Bala Nair, Ronit Katz, Jonathan Himmelfarb, Nisha Bansal, and Su-In Lee.
\newblock From local explanations to global understanding with explainable ai for trees.
\newblock \emph{Nature machine intelligence}, 2\penalty0 (1):\penalty0 56--67, 2020.

\bibitem[Biesuz et~al.(2018)Biesuz, Spiridigliozzi, Dell’Agli, Bortolotti, and Sglavo]{biesuz2018synthesis}
Mattia Biesuz, Luca Spiridigliozzi, Gianfranco Dell’Agli, Mauro Bortolotti, and Vincenzo~M Sglavo.
\newblock Synthesis and sintering of (mg, co, ni, cu, zn) o entropy-stabilized oxides obtained by wet chemical methods.
\newblock \emph{Journal of materials science}, 53\penalty0 (11):\penalty0 8074--8085, 2018.

\bibitem[Jurica and McMullen(2021)]{jurica2021automation}
Jon~A Jurica and Jonathan~P McMullen.
\newblock Automation technologies to enable data-rich experimentation: Beyond design of experiments for process modeling in late-stage process development.
\newblock \emph{Organic Process Research \& Development}, 25\penalty0 (2):\penalty0 282--291, 2021.

\bibitem[Sun and Ceder(2017)]{sun2017induction}
Wenhao Sun and Gerbrand Ceder.
\newblock Induction time of a polymorphic transformation.
\newblock \emph{CrystEngComm}, 19\penalty0 (31):\penalty0 4576--4585, 2017.

\bibitem[Mugumya et~al.(2022)Mugumya, Rasche, Rafferty, Patel, Mallick, Mou, Bobb, Gupta, and Jiang]{mugumya2022synthesis}
Jethrine~H Mugumya, Michael~L Rasche, Robert~F Rafferty, Arjun Patel, Sourav Mallick, Mingyao Mou, Julian~A Bobb, Ram~B Gupta, and Mo~Jiang.
\newblock Synthesis and theoretical modeling of suitable co-precipitation conditions for producing nmc111 cathode material for lithium-ion batteries.
\newblock \emph{Energy \& Fuels}, 36\penalty0 (19):\penalty0 12261--12270, 2022.

\bibitem[Van~Bommel and Dahn(2009)]{van2009analysis}
Andrew Van~Bommel and JR~Dahn.
\newblock Analysis of the growth mechanism of coprecipitated spherical and dense nickel, manganese, and cobalt-containing hydroxides in the presence of aqueous ammonia.
\newblock \emph{Chemistry of Materials}, 21\penalty0 (8):\penalty0 1500--1503, 2009.

\bibitem[Wang et~al.(2011)Wang, Belharouak, Koenig, Zhou, and Amine]{wang2011growth}
Dapeng Wang, Ilias Belharouak, Gary~M Koenig, Guangwen Zhou, and Khalil Amine.
\newblock Growth mechanism of ni0. 3mn0. 7co3 precursor for high capacity li-ion battery cathodes.
\newblock \emph{Journal of Materials Chemistry}, 21\penalty0 (25):\penalty0 9290--9295, 2011.

\bibitem[Burton et~al.(1951)Burton, Cabrera, and Frank]{burton1951growth}
W-K\_ Burton, N~t Cabrera, and FC~Frank.
\newblock The growth of crystals and the equilibrium structure of their surfaces.
\newblock \emph{Philosophical Transactions of the Royal Society of London. Series A, Mathematical and Physical Sciences}, 243\penalty0 (866):\penalty0 299--358, 1951.

\bibitem[Uwaha(2016)]{uwaha2016introduction}
Makio Uwaha.
\newblock Introduction to the bcf theory.
\newblock \emph{Progress in Crystal Growth and Characterization of Materials}, 62\penalty0 (2):\penalty0 58--68, 2016.

\bibitem[Duan et~al.(2014)Duan, Wang, Guo, and Gregory]{duan2014zeta}
Jinming Duan, Junsheng Wang, Tingting Guo, and John Gregory.
\newblock Zeta potentials and sizes of aluminum salt precipitates--effect of anions and organics and implications for coagulation mechanisms.
\newblock \emph{Journal of Water Process Engineering}, 4:\penalty0 224--232, 2014.

\bibitem[Adair et~al.(2001)Adair, Suvaci, and Sindel]{adair2001surface}
JH~Adair, E~Suvaci, and J~Sindel.
\newblock Surface and colloid chemistry.
\newblock 2001.

\bibitem[Serrano-Lotina et~al.(2023)Serrano-Lotina, Portela, Baeza, Alcolea-Rodr{\'\i}guez, Villarroel, and {\'A}vila]{serrano2023zeta}
A~Serrano-Lotina, Raquel Portela, P~Baeza, V{\'\i}ctor Alcolea-Rodr{\'\i}guez, M~Villarroel, and PJCT {\'A}vila.
\newblock Zeta potential as a tool for functional materials development.
\newblock \emph{Catalysis Today}, 423:\penalty0 113862, 2023.

\bibitem[Christensen et~al.(2021)Christensen, Yunker, Shiri, Zepel, Prieto, Grunert, Bork, and Hein]{christensen2021automation}
Melodie Christensen, Lars~PE Yunker, Parisa Shiri, Tara Zepel, Paloma~L Prieto, Shad Grunert, Finn Bork, and Jason~E Hein.
\newblock Automation isn't automatic.
\newblock \emph{Chemical science}, 12\penalty0 (47):\penalty0 15473--15490, 2021.

\bibitem[Kashchiev et~al.(1991)Kashchiev, Verdoes, and Van~Rosmalen]{kashchiev1991induction}
D~Kashchiev, D~Verdoes, and GM~Van~Rosmalen.
\newblock Induction time and metastability limit in new phase formation.
\newblock \emph{Journal of crystal growth}, 110\penalty0 (3):\penalty0 373--380, 1991.

\bibitem[Boser et~al.(1992)Boser, Guyon, and Vapnik]{boser1992training}
Bernhard~E Boser, Isabelle~M Guyon, and Vladimir~N Vapnik.
\newblock A training algorithm for optimal margin classifiers.
\newblock In \emph{Proceedings of the fifth annual workshop on Computational learning theory}, pages 144--152, 1992.

\bibitem[Sch{\"u}tt et~al.(2020)Sch{\"u}tt, Chmiela, von Lilienfeld, Tkatchenko, Tsuda, and M{\"u}ller]{schutt2020machine}
Kristof~T Sch{\"u}tt, Stefan Chmiela, O~Anatole von Lilienfeld, Alexandre Tkatchenko, Koji Tsuda, and Klaus-Robert M{\"u}ller.
\newblock Machine learning meets quantum physics.
\newblock \emph{Lecture Notes in Physics}, 2020.

\end{thebibliography}

\end{document}

\title{Kernel learning assisted synthesis\\ condition exploration for ternary spinel}

\author[1]{\orcidlink{0000-0003-4737-7123}~Yutong Liu\textsuperscript{$\dagger$}}
\author[2]{\orcidlink{0000-0001-5696-9193}~Mehrad Ansari\textsuperscript{$\dagger$,$\ast$}}
\author[3]{\orcidlink{0009-0004-3383-0307}~Robert Black}
\author[1,2]{\orcidlink{0000-0003-2937-3188}~Jason Hattrick-Simpers\textsuperscript{$\ast$}}

\affil[1]{\small{Department of Materials Science and Engineering, University of Toronto, Toronto, ON, Canada}}
\affil[2]{\small{Acceleration Consortium, University of Toronto, Toronto, ON, Canada}}
\affil[3]{\small{Clean Energy Innovation Research Center, National Research Council Canada, Mississauga, ON, Canada}}

\renewcommand{\thefootnote}{\fnsymbol{footnote}}
\footnotetext{$\dagger$ These authors contributed equally.}
\footnotetext{$\ast$ Correspondence to: \{mehrad.ansari, jason.hattrick.simpers\}@utoronto.ca}

\maketitle
\newpage
\begin{figure}[H]
\includegraphics[width=\textwidth, center]{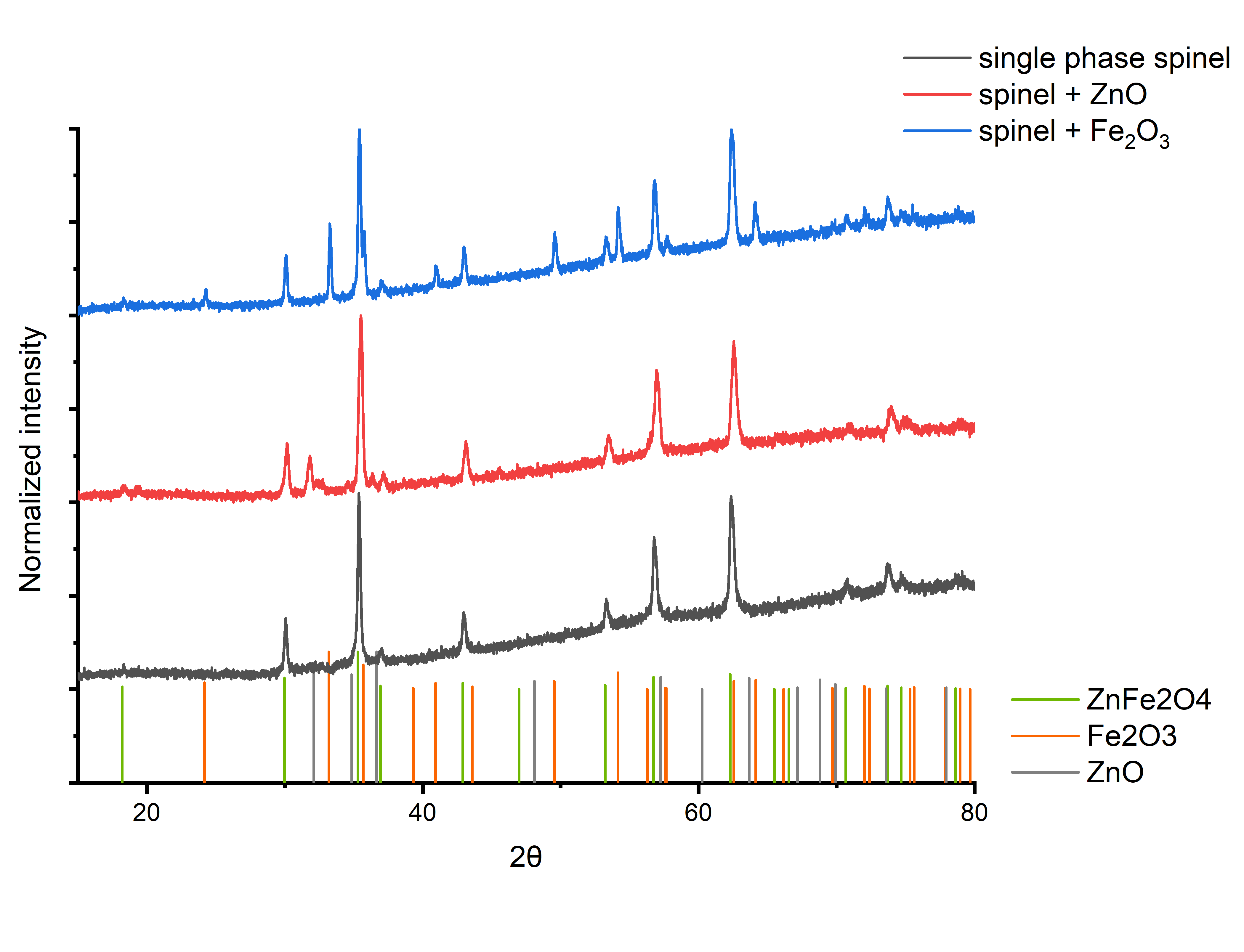}
\caption{X-ray diffraction patterns for identified single phase spinel and spinel with another secondary phase.}
\end{figure}

\begin{figure}[H]
\includegraphics[width=\textwidth, center]{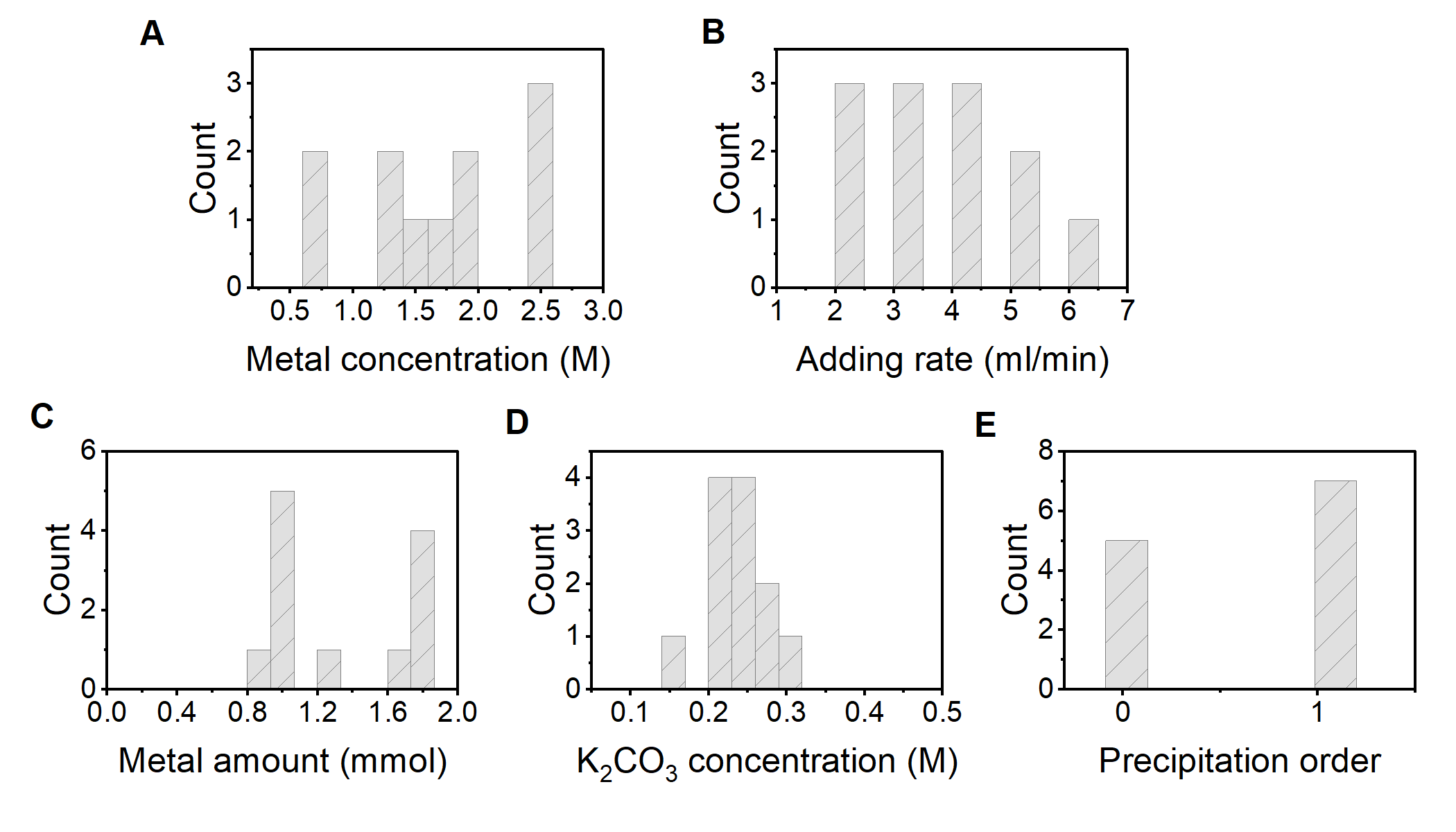}
\caption{Distributions for of five parameters used in machine
learning model in the 'missing region' for (A) metal precursor concentration, (B) adding rate, (C) metal amount, (D)
K$_2$CO$_3$ concentration and (E) precipitation order. No single phase spinel formed. }
\label{fig:workflow}
\end{figure}

\begin{figure}[H]
\includegraphics[width=\textwidth, center]{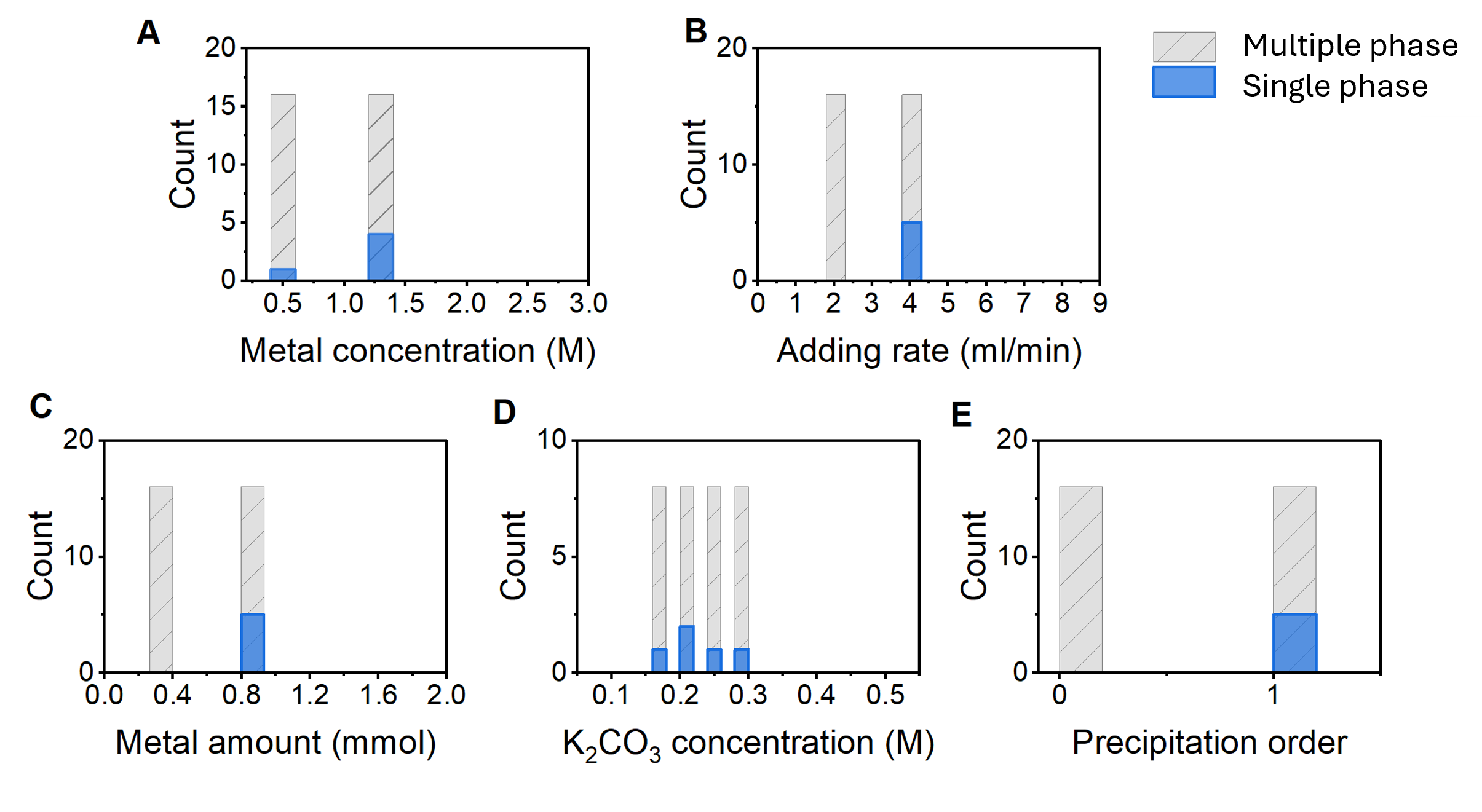}
\caption{Distributions for single phase versus multiple phase of five parameters used in machine
learning model in additional test experiments for (A) metal precursor concentration, (B) adding rate, (C) metal amount, (D)
K$_2$CO$_3$ concentration and (E) precipitation order.}
\label{fig:workflow}
\end{figure}

\begin{figure}[H]
\includegraphics[width=0.85\textwidth, center]{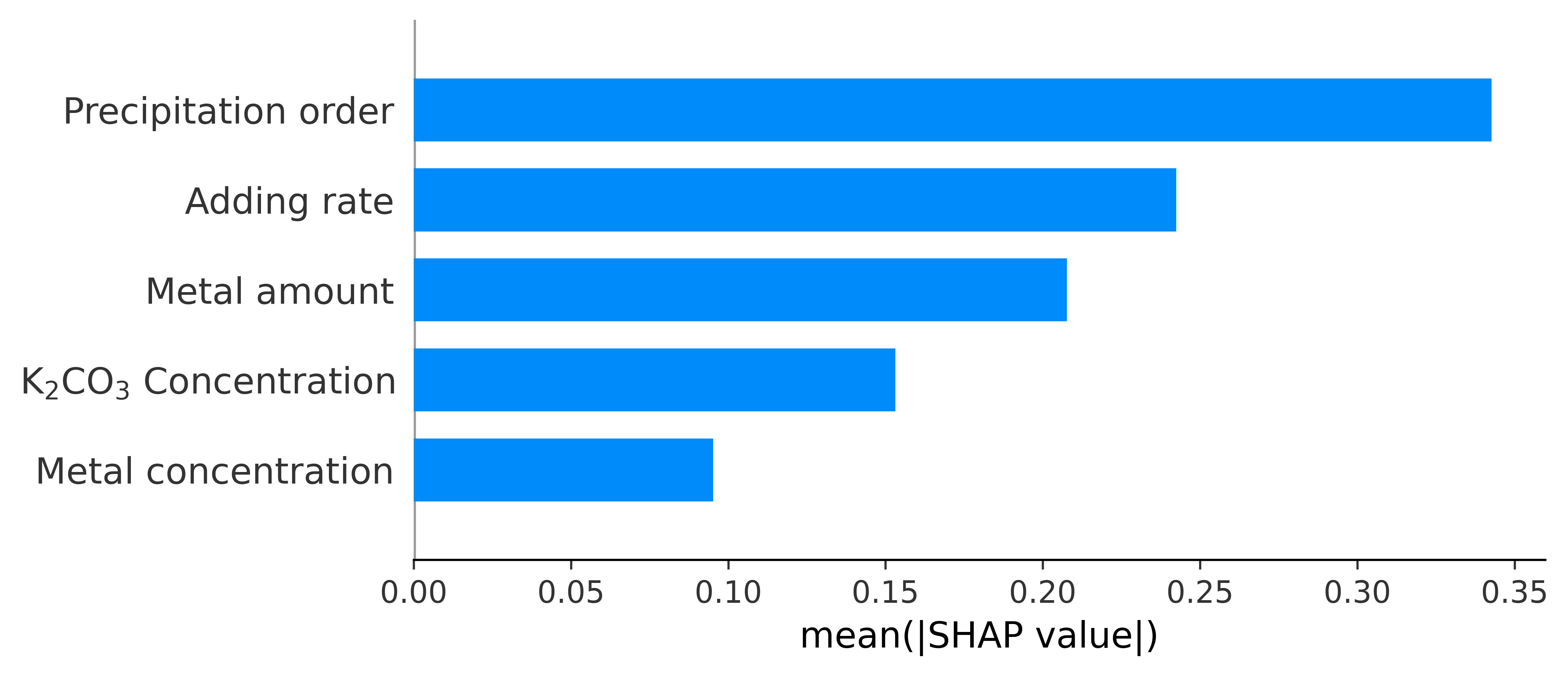}
\caption{Feature importance ranking based on SHAP values for the kernel model.}
\label{fig:workflow}
\end{figure}